\documentclass[journal]{IEEEtran}

\title{Design of Block Transceivers with Decision Feedback Detection}
\author{Fang Xu, Timothy~N.~Davidson, 
        Jian-Kang Zhang, and K.~Max~Wong
\thanks{This work was supported in part by the National Science and
   Engineering Research Council of Canada.
The work of the second author is also supported by the
Canada Research Chairs program.}%
\thanks{The authors are with the 
 Department of Electrical and Computer
Engineering, McMaster University, 
Hamilton, Ontario, L8S 4K1, Canada.
Email: xufang03@hotmail.com, \{davidson,wongkm\}@mcmaster.ca, jkzhang@mail.ece.mcmaster.ca}}

\usepackage{cite}
\usepackage{amsmath}
\usepackage{latexsym}
\usepackage{amssymb}
\usepackage{graphicx}
\usepackage{fix2col}
\usepackage{url}

\def\erfc{{\operatorname{erfc}}}
\def\tr{{\operatorname{tr}}}

\newtheorem{lemma}{Lemma}
\newtheorem{proposition}{Proposition}

\newcommand{\bA}{{\mathbf A}}
\newcommand{\bB}{{\mathbf B}}
\newcommand{\bC}{{\mathbf C}}
\newcommand{\bD}{{\mathbf D}}

\newcommand{\bF}{{\mathbf F}}

\newcommand{\bH}{{\mathbf H}}
\newcommand{\bI}{{\mathbf I}}

\newcommand{\bP}{{\mathbf P}}
\newcommand{\bQ}{{\mathbf Q}}
\newcommand{\bR}{{\mathbf R}}
\newcommand{\bS}{{\mathbf S}}

\newcommand{\bU}{{\mathbf U}}
\newcommand{\bV}{{\mathbf V}}
\newcommand{\bW}{{\mathbf W}}
\newcommand{\bX}{{\mathbf X}}

\newcommand{\bZ}{{\mathbf Z}}
\newcommand{\bZero}{{\mathbf 0}}

\newcommand{\be}{{\mathbf e}}

\newcommand{\bs}{{\mathbf s}}

\newcommand{\bu}{{\mathbf u}}
\newcommand{\bv}{{\mathbf v}}

\newcommand{\by}{{\mathbf y}}
\newcommand{\bz}{{\mathbf z}}

\newcommand{\bGamma}{\boldsymbol{\Gamma}}
\newcommand{\bLambda}{\boldsymbol{\Lambda}}
\newcommand{\bTheta}{\boldsymbol{\Theta}}
\newcommand{\bPhi}{\boldsymbol{\Phi}}
\newcommand{\bPsi}{\boldsymbol{\Psi}}

\newcommand{\zf}{\text{\textsc{zf}}}
\newcommand{\mmse}{\text{\textsc{mmse}}}

\begin{document}
\maketitle

\begin{abstract}
This paper presents a method for jointly designing  
the transmitter-receiver pair in a block-by-block
communication system that employs (intra-block) decision feedback detection. 
We provide closed-form expressions for transmitter-receiver pairs that
simultaneously minimize the arithmetic mean 
squared error (MSE) at the decision point 
(assuming perfect feedback), the geometric MSE, and the
bit error rate of a uniformly bit-loaded system at moderate-to-high 
signal-to-noise ratios.
Separate expressions apply for the ``zero-forcing'' and
``minimum MSE'' (MMSE) decision feedback structures. 
In the MMSE case, the proposed design also maximizes the Gaussian
mutual information and suggests that 
one can approach the capacity of the block transmission system using 
(independent instances of) the
same (Gaussian) code for each element of the block.
Our simulation studies indicate that the proposed 
 transceivers perform significantly better than standard 
transceivers, and that they retain their performance advantages 
in the presence of error propagation.
\end{abstract}


\section{Introduction}

Block-by-block communication is an effective scheme for the transmission 
of data over dispersive media;
e.g.,~\cite{Kasturia,Lechleider,kaleh,scaglione,scaglione1}. 
In such ``vector'' communication schemes, 
blocks of data are transmitted in a manner that avoids
interference between the received blocks, and hence
the detector need only operate on a block-by-block basis.
Two popular examples of block-by-block communication schemes 
are orthogonal frequency division 
multiplexing (OFDM)~\cite{bingham} and discrete multi-tone 
modulation (DMT)~\cite{chow}. 
In addition, certain multiple antenna 
systems operate in a block-by-block fashion
(e.g.,~\cite{foschini,ginis,scaglione2002,palomar,Tarokh,HHLD}),
and block-by-block detection schemes appear in some multiuser
detectors for  synchronous CDMA systems~\cite{Duel-Hallen-Synch,Duel-Hallen,varanasi}.
In general, an optimal detector for a block transmission system must make 
a decision on the received data block as a whole,
although in  certain cases, such as OFDM and DMT, the elements of that block 
can be decoupled and  simpler detection schemes obtained. 
Unfortunately, maximum likelihood detection of the transmitted 
vector can be rather computationally expensive, and simpler detectors 
based on linear equalization and (disjoint) symbol-by-symbol 
detection may incur a significant performance loss. 
A useful compromise between performance and complexity can be obtained 
by employing  intra-block decision feedback
detection~\cite{kaleh,foschini,cioffi-forney-GDFE,ginis,biglieri,stamoulis,xu,guess,Duel-Hallen-Synch,Duel-Hallen,varanasi}.
In an intra-block decision feedback detector the individual
symbols which constitute a 
given block are detected sequentially, 
with the ``intra-block interference'' from previously detected symbols 
being subtracted before the decision on the current symbol is made. 
Such schemes fall into the class of generalized decision feedback
equalizers~\cite{cioffi-forney-GDFE}.
In multiple antenna communication schemes intra-block decision feedback 
is sometimes referred to as ``nulling and cancelling''~\cite{foschini,biglieri,ginis},
and in multi-user detection the corresponding concept is
sometimes referred to as ``successive interference 
cancellation''~\cite{Duel-Hallen,Duel-Hallen-Synch,varanasi,guess}. 

The goal of the present paper is to jointly design the linear 
transmitter matrix and the receiver feedforward and feedback matrices 
so as to optimize the performance
of a block-by-block communication system with 
an intra-block decision feedback detector (BDFD). 
 The design is based on knowledge of the channel, and hence
is an appropriate choice for systems in which there is  
timely, reliable feedback from the receiver to the transmitter. 
The proposed approach provides closed-form  expressions for 
transceivers that minimize the arithmetic mean (over the block)
of the expected squared errors (MSE)  
at the input to the (scalar) decision device that is implicit in the BDFD,
under the standard
assumption~\cite{salz,falconer-foschini,Belfiore_Park_79,cioffi,Yang,cioffi-forney-GDFE} 
that the previous decisions were correct.
The expressions depend on the nature of the BDFD, and separate
expressions are provided for the zero-forcing (ZF) and 
minimum mean square error (MMSE) BDFDs.
In order to help distinguish our designs from previous work, we point
out that if one is given a transmitter matrix, the design of the feedforward
and feedback matrices of a ZF or MMSE-BDFD that minimize the MSE
is well known; e.g.,~\cite{BarryLeeMesserschmitt,cioffi,cioffi-forney-GDFE,salz,falconer-foschini,biglieri,ginis}.
However, the joint minimum MSE design of the transmitter and receiver matrices
has previously been deemed to be
difficult (e.g.,~\cite[p.~1338]{Yang}), 
and hence several authors have suggested minimizing 
a particular lower bound on the MSE, namely the  
geometric  mean of the expected squared errors; e.g., \cite{cioffi,Yang,cioffi-forney-GDFE}. 
We will  minimize the geometric MSE as the first step
in our approach, but we will also show how  
the unitary matrix that parameterizes  the set of
transceivers which minimize the geometric MSE can be chosen so that
the (arithmetic) MSE attains its minimized lower bound.

Transceivers designed in the manner we propose have several additional 
desirable properties. In particular, the inputs to the
(scalar) decision device are uncorrelated and have 
equal  signal-to-interference-and-noise ratios (SINRs).
In fact, the minimum SINR over the elements of the block is
maximized. As a result, 
the average bit error rate (BER) is (essentially) minimized.
More precisely, for systems with a ZF-BDFD
our design minimizes the average BER for (uncoded) uniform 
QPSK signalling at moderate-to-high
signal-to-noise ratios (SNRs), and 
also minimizes the dominant components of the
BER for uniform 
$\mathcal{M}$-ary QAM signalling.%
\footnote{Our design for the ZF-BDFD
coincides with the one that 
minimizes the block error rate~\cite{zhang,zhang1}, but the design
approach taken in the present paper is substantially different from
that taken in~\cite{zhang,zhang1}.} 
For systems with an MMSE-BDFD, our design minimizes the average BER
under an assumption that
the residual intra-block interference is Gaussian.

For the MMSE-BDFD, it is reasonably well 
known~\cite{salz,falconer-foschini,cioffi,Yang,cioffi-forney-GDFE}
that any transmitter that minimizes the geometric MSE 
(including the proposed design) also
maximizes the mutual information between the transmitter and 
receiver for Gaussian signals. 
However, the standard choice from
the set of transmitters that minimize the geometric MSE 
does not minimize the (arithmetic) MSE and 
produces inputs to the decision device that have
potentially different SINRs for each element of the block. 
Therefore, in order to
achieve reliable communication at rates which approach the
capacity of the block transmission system, different codes
(and constellations) may need to be applied for each
element of the block~\cite{cioffi-forney-GDFE}.
An advantage of the  proposed design is that 
from within the set of transmitters that minimize the
geometric MSE (and maximize the Gaussian mutual information), 
we obtain a transceiver that also minimizes the arithmetic MSE,
minimizes the  BER,  and provides uncorrelated
inputs to the decision device that have identical (and maximized) SINRs.
Since the MMSE-BDFD is a ``canonical'' 
receiver~\cite{cioffi,cioffi-forney-GDFE,Guess_Varanasi_MMSE_DFE}, this
 suggests that by using the proposed design, reliable communication
at rates approaching the capacity of the block transmission 
system can be achieved  by using independent instances of the same 
(Gaussian) code in each element of block.

As mentioned earlier, our designs are based on the standard
assumption~\cite{salz,falconer-foschini,Belfiore_Park_79,cioffi,Yang,cioffi-forney-GDFE} 
that the previous symbols were correctly detected. 
However, error propagation is not catastrophic in block-by-block
communication schemes because errors can only propagate
within a single block (e.g.,~\cite{cioffi-forney-GDFE}
and Section~\ref{sec:modelling}). Bounds for the
conventional symbol-by-symbol decision feedback equalizer 
(DFE)~\cite{dutt,Beaulieu}
also suggest that good performance should be maintained in the presence
of error propagation, and our simulations confirm this
prediction. Furthermore, our
 simulation studies indicate that the proposed
 transceivers perform significantly better than standard
transceivers, and that they retain their performance advantages
in the presence of error propagation.

{\textit{Notation:}} The notation adopted in this paper is fairly standard. 
We conform to the following conventions: scalars are denoted by lower case 
letters; vectors by bold lower case letters;
and  matrices by bold upper case letters.
 The symbol ${\bf I}_{N}$ denotes the identity matrix of size $N$,
and $\boldsymbol{0}_{N\times M}$ denotes the $N\times M$ matrix of zeros.
 The symbol $|{\bf A}|$ denotes the determinant of a matrix ${\bf A}$,
and  $\operatorname{tr}({\bf A})$ denotes its trace. The symbol
 $E[\cdot]$ denotes the expectation operator;
 $(\cdot)^{H}$ the complex-conjugate transpose operation;
 $(\cdot)^T$ the transpose operation; 
and $[\cdot]_{ij}$ denotes the element at the intersection
of the $i$th row and $j$th column of a matrix.

\section{Block-by-block Transmission}
\label{sec:modelling}
\begin{figure*}
\begin{center}
\resizebox{.7\textwidth}{!}{\includegraphics{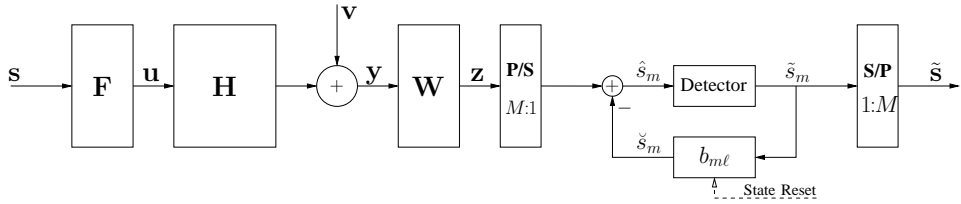}}
\caption{A generic block-by-block communication system
with intra-block decision feedback detection. 
The  {\bf{P/S}} block denotes parallel-to-serial conversion with the last
element of the input block becoming the first output,
and the {\bf{S/P}} block denotes serial-to-parallel conversion with the first
input becoming the last element of the output block.}
\label{fig:genericblockTX_DFE}
\end{center}
\end{figure*}
  
We consider the generic block-by-block transmission 
system with intra-block decision feedback detection
illustrated in Fig.~\ref{fig:genericblockTX_DFE}.
In this system, a block of  $M$ data symbols, $\mathbf{s}$, is
linearly precoded to construct a block of $K\geq M$
channel symbols, $\bu = \bF \bs$,
which is  transmitted over the channel.
 The receiver independently processes  a block of $P\geq M$
received samples in order to detect the data vector~$\bs$.
The received block, $\by$, can be written as
\begin{equation}
    \by = {\bf{HF}}\bs + \bv, \label{equalizedoutputsim}
\end{equation}
where the $P\times K$ matrix $\mathbf{H}$ captures the effects
of the channel, and $\bv$ is a length $P$ vector of additive
noise samples. We will assume that the noise is  
circularly symmetric~\cite{picinbono} (or, proper~\cite{NeeserMassey})
and Gaussian, with zero mean and positive definite correlation matrix
$E[\bv\bv^H]=\mathbf{R}_{vv}$.
We will also assume that the data symbols have zero mean and are white,%
\footnote{In the case where $E[\mathbf{ss}^H]$ is not a scaled identity matrix,
a data whitening matrix can readily be absorbed into the precoder, so long as
the data covariance matrix is known (and full rank).}
 of unit energy, and not correlated with the noise,
(i.e., $E[\mathbf{ss}^H]=\mathbf{I}$ and $E[\mathbf{sv}^H]=0$).
The model in \eqref{equalizedoutputsim} is applicable in many applications,
including zero-padded or cyclic-prefixed block transmission over a scalar 
finite impulse response channel 
that is constant over the duration of the block;
e.g.,~\cite{chan,dingy,scaglione1,scaglione,Kasturia,kaleh,Lechleider,stamoulis}.
In the zero-padded case $\mathbf{H}$ is a tall, lower triangular, full column rank 
Toeplitz matrix whose columns contain the impulse response of the channel,
and in the cyclic-prefixed case $\mathbf{H}$ is a square circulant matrix
whose columns contain the channel impulse response. 
The model in \eqref{equalizedoutputsim} is also applicable in: 
vector transmission over a narrowband multiple antenna channel 
(e.g., \cite{foschini,ginis}), in which case $\mathbf{H}$ 
has no deterministic structure;
in space-time block transmission over a (quasi-static)
narrowband multiple antenna channel
(e.g., \cite{Tarokh,HHLD}), in which case $\mathbf{H}$
has a block diagonal structure;
and in block transmission over a (quasi-static)
 frequency-selective multiple antenna
channel (e.g.,~\cite{scaglione2002,palomar}), in which case
$\mathbf{H}$ is either block Toeplitz or block circulant.

The intra-block decision feedback detector 
first pre-processes the received block $\by$ with an
 $M \times P$ feedforward matrix $\bW$
to form $\bz={\bW}\by$. (The functional form of
$\bW$ depends on whether the ZF- or MMSE-BDFD is implemented;
see Section~\ref{sec:deriv}.)
 The detection of the transmitted symbols $s_m=[\mathbf{s}]_m$
then proceeds sequentially, starting from $m=M$, by making a scalar decision on
$\hat{s}_M = z_M$ and then $\hat{s}_m = z_m - \breve{s}_m$,
$m=M-1, M-2,\dots,1$, where $\breve{s}_m=\sum_{\ell=m+1}^M b_{m\ell} \tilde{s}_\ell$
is the output of the feedback filter, with $b_{m\ell}$ being its coefficients. 
The states of that filter, $\tilde{s}_\ell$,  are the previously 
detected symbols in the block and the filter coefficients are different 
for each element of the block (indexed by $m$). 
Once a given block has been detected, the states of the feedback filter are reset to zero. 
That is, the symbols are detected on a block-by-block basis and hence 
error propagation between blocks is avoided.

\begin{figure}
   \centering 
   \resizebox{0.48\textwidth}{!}{\includegraphics{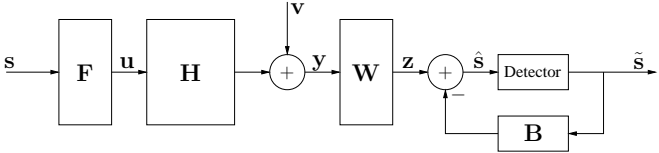}}
   \caption{A convenient conceptual model for Fig.~\ref{fig:genericblockTX_DFE}.}
  \label{fig:model}
\end{figure}

If the filter coefficients $b_{m\ell}$ are arranged
in a strictly upper triangular $M\times M$ matrix 
\begin{eqnarray*}
 \bB=\left[\begin{array}{ccccc}
    0 & b_{12} & b_{13} & \cdots & b_{1M}\\
    0 & 0      & b_{23} & \cdots & b_{2M}\\
    \vdots & \ddots & \ddots & \ddots & \vdots \\
    0 & 0      & 0      &  \cdots & b_{(M-1)M}\\
    0 & 0      & 0      &  \cdots & 0 \end{array}\right],
\end{eqnarray*}
the operation of the block transceiver in Fig.~\ref{fig:genericblockTX_DFE}
is equivalent to successively making decisions on the elements of 
      \begin{equation}\label{equz}
         \hat{\bs} = {\bf WHFs}+{\bf Wv}-\bB\tilde{\bs},
      \end{equation}
starting from the $M$th row. That interpretation leads to
the convenient conceptual model in Fig.~\ref{fig:model}.
We observe that when $\bB=\mathbf{0}$, the
system in Fig.~\ref{fig:model} reduces to a block transmission
system with linear equalization and disjoint detection; 
e.g.,~\cite{chan,dingy,scaglione1,scaglione,scaglione2002,palomar}. 
In fact, many of the results for the linear case can be obtained
by setting $\bB=\mathbf{0}$ in the expressions we will derive  herein.

If we denote the error between the input to the detector and the transmitted 
data symbols by $\be=\hat{\bs}-\bs$, then
\begin{equation}\label{error}
    \be  =  (\mathbf{WHF}-\bI)\bs - \bB \tilde{\bs} + \mathbf{Wv}.
\end{equation}
Under the assumption of correct past decisions (i.e., when deciding $s_m$, 
$\tilde{s}_\ell = s_\ell$ for all $m+1\leq \ell \leq M$),
 $\be$ simplifies to 
\begin{equation}\label{error0}
    \be  =  (\bf{WHF}-\bI - \bB)\bs + \bf{Wv}.
 \end{equation}
The covariance of this error will play a key role in our designs. 
Under our statistical models for $\bs$ and $\bv$,
the covariance matrix of the error is
\begin{multline}\label{Reet}
  \mathbf{R}_{ee} =  E[{\bf{ee}}^H]
    =  (\mathbf{WHF}-\mathbf{B}-\mathbf{I})(\mathbf{WHF}-\mathbf{B}-\mathbf{I})^H \\ +\mathbf{WR}_{vv}\mathbf{W}^H.
\end{multline}
The (arithmetic) MSE of the detector input is simply
${\bar{e}}^2=\operatorname{tr}\Bigl(E[(\hat{\bf s}-{\bf{s}})(\hat{\bf
s}-{\bf{s}})^H)]\Bigr)/M=\operatorname{tr}(\mathbf{R}_{ee})/M$.

\section{Minimum MSE Transceivers}
\label{sec:deriv}
In this section, our goal is to jointly design the
transceiver elements $\bF$, $\bB$, and $\bW$ so that the 
(arithmetic) MSE is minimized, subject to a bound, $p_0$, on the average
transmitted power, and constraints which ensure that the receiver performs 
either ZF or MMSE decision-feedback detection.
 The average transmitted power is given by 
$E\bigl[\operatorname{tr}\bigl(\mathbf{F}\mathbf{s}(\mathbf{F}\mathbf{s})^H\bigl)\bigr]
=\operatorname{tr}(\mathbf{FF}^H)$,
and hence the design problem can be stated as
\begin{subequations}\label{opt1}
\begin{align}
\min_{{\bf F},{\bf B},{\bf W}} & \ {\operatorname{tr}}\Bigl((\mathbf{WHF}-\mathbf{B}-\mathbf{I})
(\mathbf{WHF}-\mathbf{B}-\mathbf{I})^H\Bigr. \notag \\
&\qquad\qquad\qquad\qquad\qquad\qquad
\Bigl.+\mathbf{WR}_{vv}\mathbf{W}^H\Bigr) \\
\mbox{subject to}  & \ \   {\operatorname{tr}}({\bf FF}^{H}) \leq p_0, 
\ \text{and}\\
 & \ \ \text{a functional relationship between $\bF$, $\bB$ and $\bW$.}
\end{align}
\end{subequations}
The functional relationship between $\bF$, $\bB$ and $\bW$ determines
whether the BDFD is of the ZF type or the MMSE type.
This optimization problem is rather difficult to solve directly
because it is not convex, and hence is subject to
the standard difficulties associated with the potential for
multiple local minima. However, we will use the following stages
to find a solution $({\bf F},{\bf B},{\bf W})$ whose
performance is optimal:
\begin{enumerate}
  \item Obtain a (tight) lower bound on the MSE, and minimize
  that lower bound, subject to the constraint on transmission
  power.
  \item Derive a triple $({\bf F},{\bf B},{\bf W})$ whose performance achieves the minimized lower bound.
\end{enumerate}
In the following subsections, we will perform the above stages to
 obtain the minimized
lower bounds on the MSE and optimal transceivers for the ZF
and MMSE BDFDs, respectively.

The matrix $\bH^H\bR_{vv}^{-1}\bH$ will  play a key role in our designs. For later
convenience we let
\begin{equation}
\bV\bLambda\bV^H = \bH^H\bR_{vv}^{-1}\bH
\label{eq:eigHRH}
\end{equation}
represent the eigenvalue decomposition of $\bH^H\bR_{vv}^{-1}\bH$, with eigenvalues $\lambda_i$
 arranged in non-increasing order along the diagonal of $\bLambda$. For an integer $1\leq k\leq K$, we
 also define $\tilde{\bV}_k$ to be the first $k$ columns of $\bV$ and 
 $\tilde{\bLambda}_k$ to be the upper left $k\times k$ block of $\bLambda$. 
 In the development of our designs, we will find it convenient  to parameterize the
  $K\times M$ precoder matrix  $\bF$ of rank~$q$ in terms of its singular value decomposition,
 \begin{equation}
 \bF = \bTheta \begin{bmatrix} \bPhi & \bZero_{q\times (M-q)}\end{bmatrix}
 \bPsi
 \label{eq:svdF}
 \end{equation}
 where $\bTheta$ contains $q$~columns of a $K\times K$ unitary matrix,
 $\bPhi$ is a diagonal positive definite $q\times q$ matrix, and $\bPsi$ is an
 $M\times M$ unitary matrix.

\subsection{Zero-forcing BDFD}
\label{sec:deriv_ZF}
The zero-forcing criterion imposes the following relationship between
$\mathbf{W}, \mathbf{F}$ and $\mathbf{B}$ (see (\ref{error0})):
\begin{equation}\label{equzf}
            {\mathbf{WHF}}={\mathbf{B}}+{\mathbf{I}}.
         \end{equation}
Given a $P\times K$ matrix $\bH$
and an integer $M\leq \min\{P,K\}$, there exists a
$K\times M$ matrix $\bF$, an $M\times P$ matrix $\bW$ and
an $M\times M$ strictly upper triangular matrix $\bB$
such that \eqref{equzf} is satisfied if and only if 
$\operatorname{rank}(\bH)\geq M$, and we will make
the assumption that this condition holds.%
\footnote{If $M$ were a design variable, rather than a parameter of
the problem, one could guarantee that this condition holds by
simply choosing $M\leq \operatorname{rank}(\bH)$.}
 In order to satisfy \eqref{equzf}, $\bF$ must be
 chosen so that it has rank $M$ and that 
$\operatorname{rank}(\bH\bF) = M$.

By substituting (\ref{equzf}) into (\ref{error0}) and (\ref{Reet}), 
the covariance matrix of the error can be written as 
\begin{equation}\label{Reet1}
 \bR_{ee,\zf}=\bW\bR_{vv}\bW^H.
\end{equation} 
If we define $\breve{\bW}=\bW\bR_{vv}^{1/2}$, then
the design problem (\ref{opt1}) can be re-written as 
\begin{subequations}\label{MSEOB_Wbreve}
\begin{align}
\min_{\breve{\bW}, \bB, \bF} & \quad  
\operatorname{tr}\bigl(\breve{\bW}\breve{\bW}^H\bigr)
\\
{\mbox{subject to}} & \quad  {\operatorname{tr}}({\bf FF}^{H}) \leq{ p_0 }
\label{pow_Wbreve},  \\
    & \quad\breve{\bW}\breve{\bH}\bF= \bB+\bI,
\label{zf_Wbreve}
\end{align}
\end{subequations}
where $\breve{\bH}=\bR_{vv}^{-1/2}\bH$. 
From \eqref{MSEOB_Wbreve} it is clear that for a given $\bF$ 
for which there exists a solution to \eqref{zf_Wbreve}
and a given $\mathbf{B}$, the optimal
$\breve{\bW}$ is $\breve{\bW}=(\bB+\bI)(\breve{\bH}\bF)^+$,
where $(\cdot)^+$ denotes the (minimum-norm) Moore-Penrose
pseudo-inverse. Therefore, the optimal receiver feedforward
matrix can be written as 
\begin{equation}
\label{equzf1}
\bW_{\zf}= (\bB+\bI)(\breve{\bH}\bF)^+\bR_{vv}^{-1/2}.
\end{equation}
Since $\breve{\bH}\bF$ has at least as
many rows as it has columns and   has
full column rank, 
\begin{equation}(\breve{\bH}\bF)^+ =
(\bF^H\breve{\bH}^H\breve{\bH}\bF)^{-1}
\bF^H\breve{\bH}^H.
\label{eq:ps_inv}
\end{equation}
If we let $\bU=\bB+\bI$, the  design problem in 
\eqref{MSEOB_Wbreve} has been reduced to
\begin{subequations}\label{MSEOB}
\begin{align}
\min_{\bf{U,F}} & \quad  
{\operatorname{tr}}\left({\bf U}
(\breve{\bH}\bF)^{+}
\bigl((\breve{\bH}\bF)^{+}\bigr)^H{\bf U}^{H}\right) 
\label{MSEOB_obj}
\\
{\mbox{subject to}} & \quad  {\operatorname{tr}}({\bf FF}^{H}) \leq{ p_0 }\label{pow},  \\
          & \quad {\bf U}\, {\mbox{being 
a unit-diagonal upper-triangular matrix.}}\label{up}
\end{align}
\end{subequations}

The first stage in our solution of \eqref{MSEOB}
is to derive and minimize a lower bound on the objective function (\ref{MSEOB_obj}).
The lower bound that we will use is a simple consequence of the
arithmetic-geometric mean inequality~\cite[p.~535]{horn}. In particular,
for an $M\times M$ positive semidefinite matrix $\bX$,
\begin{equation} \label{eq:AGMI}
\operatorname{tr}(\bX)/ M \geq |\bX|^{1/M},
\end{equation}
with equality holding if and only if $\bX=\alpha\bI$ for some $\alpha\geq0$.
For convenience, we will refer to \eqref{eq:AGMI} as the trace-determinant
inequality.

Applying \eqref{eq:AGMI} to (\ref{MSEOB_obj}), 
a lower bound on the mean-square error is 
\begin{subequations}\label{LOWMSE}
\begin{align}
{\bar{e}}_{\zf}^2  
= \operatorname{tr}(\bW_{\zf}\bR_{vv}\bW_{\zf}^H)/M
&\geq \Bigl|\bU
(\breve{\bH}\bF)^+\bigl((\breve{\bH}\bF)^+\bigr)^H
\bU^H\Bigr|^{1/M}
\label{LOWMSE2a}
\\
& = |\bF^H\breve{\bH}^H\breve{\bH}\bF|^{-1/M},
\label{LOWMSE2}
\end{align}
\end{subequations}
where we have used the fact that $\bU$ is a unit-diagonal 
upper-triangular matrix and thus $|\bU|=1$,
and the expression for $(\breve{\bH}\bF)^+$ in \eqref{eq:ps_inv}.
Observe that (\ref{LOWMSE2}) depends only
on the transmitter $\bF$ and is  independent of $\bf{U}=\bB+\bI$. 
It is also of interest to point out that the
bound in \eqref{LOWMSE2a} is equivalent to stating
that the arithmetic MSE is bounded below by the
geometric MSE; i.e. $\operatorname{tr}(\bR_{ee,\zf})/M
\geq |\bR_{ee,\zf}|^{1/M}$. Therefore,
the problem of minimizing the lower bound in \eqref{LOWMSE2a}
corresponds to minimizing the geometric MSE.

The lower bound in \eqref{LOWMSE} can be
minimized simply by maximizing $|\bF^H\breve{\bH}^H\breve{\bH}\bF|$; i.e.,
by solving
\begin{subequations}\label{LOWMSEF1} 
\begin{align}
\max_{{\bf F}} & \quad |\bF^H\bH^H\bR_{vv}^{-1}\bH\bF|\\
\mbox{subject to}  & \quad  {\operatorname{tr}}({\bf FF}^{H}) \leq p_0.     
\label{eq:first_power_bound}
\end{align}
\end{subequations}
Using the ordered eigen-decomposition of $\bH^H\bR_{vv}^{-1}\bH$ in
\eqref{eq:eigHRH}, and applying the trace-determinant inequality
\eqref{eq:AGMI}, 
we have that
\begin{subequations}
\begin{align}
|\bF^H\bH^H\bR_{vv}^{-1}\bH\bF| &=
|\bPhi^2|\: |\bTheta\bH^H\bR_{vv}^{-1}\bH\bTheta^H| \\
&\leq \left(\frac{\operatorname{tr}(\bPhi^2)}{M}\right)^M \prod_{i=1}^M \lambda_i 
\label{eq:invGMSEbound_a}\\
& \leq\left(\frac{p_0}{M}\right)^M \prod_{i=1}^M \lambda_i.
\label{eq:invGMSEbound}
\end{align}
\end{subequations}
Therefore, for any ZF-BDFD system, the (arithmetic) MSE is bounded below
by 
\begin{equation}
\bar{e}_{\zf}^2 \geq 
\frac{M}{p_0} \Bigl(\prod_{i=1}^M\lambda_i\Bigr)^{-1/M}.
\label{eq:zf_mse_bound}
\end{equation}
 This bound depends only on the parameters $M$ and $p_0$, and the
 $M$ largest eigenvalues of $\bH^{H}\bR_{vv}^{-1}\bH$.

The second stage of the derivation of the proposed design is to
determine matrices $\bF$ and $\bB$ so that the minimized lower
bound on the arithmetic MSE in \eqref{eq:zf_mse_bound} is achieved,
To do so, we point out that according to the trace-determinant 
inequality \eqref{eq:AGMI} and the eigenvalue decomposition of
$\bH^H\bR_{vv}^{-1}\bH$ in \eqref{eq:eigHRH}, the bound in
 \eqref{eq:invGMSEbound_a} holds with equality if and
only if $\bPhi=\alpha\bI$ for some $\alpha>0$ and 
$\bTheta=\tilde{\bV}_M\bP$, where $\tilde{\bV}_M$ was 
defined after \eqref{eq:eigHRH} and $\bP$ is an arbitrary permutation
matrix. According to the power constraint in \eqref{eq:first_power_bound},
the bound in \eqref{eq:invGMSEbound} is achieved if and only if 
$\alpha = \sqrt{p_0/M}$. Therefore,  precoders of the form
$\bF=\sqrt{p_0/M}\:\tilde{\bV}_M 
\bPsi$, where $\bPsi$ is an arbitrary $M\times M$
unitary matrix, minimize the geometric MSE of a ZF-BDFD system.
The remaining task is to determine matrices $\bPsi$
such that the bound in \eqref{LOWMSE2a} 
holds with equality. To do so, we observe that 
the trace-determinant inequality \eqref{eq:AGMI} holds with
equality if and only if  $\bX=\alpha\bI$ for some $\alpha\geq0$.
Therefore, \eqref{LOWMSE2a} holds with equality if and only if
 we can choose $\bPsi$ such that $\bR_{ee,\zf} = \sigma_e^2\bI$,
where $\sigma_e^2 = (M/p_0) \bigl(\prod_{i=1}^M \lambda_i\bigr)^{-1/M}$.
That is, we can achieve the minimized lower bound on the arithmetic
MSE if and only if we can find a $\bPsi$ such that
\begin{equation}
  \frac{M}{p_0}\bU\boldsymbol{\Psi}^{H}\tilde{\bLambda}_M^{-1}\boldsymbol{\Psi}\bU^H=\sigma_e^2\bI,
  \label{MSE3a}
\end{equation}
where $\tilde{\bLambda}_M$  was defined after \eqref{eq:eigHRH}.
By taking the Cholesky factor, solving \eqref{MSE3a} is equivalent to solving
\begin{equation}\label{QR1}
  \sqrt{\frac{M}{p_0}}\:{\bf U}\boldsymbol{\Psi}^{H}
\tilde{\bLambda}_M^{-1/2}=\sigma_e\bQ^H,
\end{equation}
where $\bQ$ is an $M\times M$ unitary matrix.
That is, we can reduce the search for a pair $(\bF,\bB)$ such that the minimized lower bound
on the MSE is achieved to the search for 
a unit-diagonal upper-triangular matrix $\bU$, and unitary matrices $\bPsi$ and $\bQ$ 
that satisfy~\eqref{QR1}.  Substituting $\sigma_e$ into (\ref{QR1}), we get:
\begin{equation}\label{QR3}
   \tilde{\bLambda}_M^{1/2}\bPsi=\bQ\bar{\bU},
\end{equation}
where $\bar{\bU}=\bigl(\prod_{i=1}^M\lambda_i\bigr)^{1/(2M)}\:\bU$. 
The following result, which is a special case of a more general result in \cite{zhang,zhang1}, 
indicates that a solution to (\ref{QR3}) exists.
\begin{lemma}
\label{lem:JKZ}
Let $\bGamma$ be a diagonal non-singular $M \times M$ matrix.
There exists a unitary matrix $\bS$ such that $\bGamma\bS$
has an equal-diagonal `R-factor' in its (standard) QR~decomposition; i.e. 
$\exists \bS$ such that $\bGamma\bS=\bQ\bR$,
where $\bQ$ is an $M \times M$ unitary matrix and $\bR$ is 
an upper-triangular matrix with equal diagonal elements 
$[\bR]_{ii}=\Bigl(\prod_{k=1}^M \gamma_{k}\Bigr)^{1/M}$ for $i=1,2,\cdots,M$,
where $\gamma_k$ is the $k$th diagonal element of $\bGamma$. \hfill$\Box$
\end{lemma}

The matrix $\bS$ in Lemma~\ref{lem:JKZ}
 can be obtained by suitably modifying Algorithm~5 in
 \cite{zhang1}. The modified algorithm is provided in 
Appendix~\ref{sec:app_JKZ}. Using that algorithm,  we can
obtain $\bPsi$ in (\ref{QR3}).
By performing the QR~decomposition of 
$\tilde{\bLambda}_M^{1/2}\boldsymbol{\Psi}$,
we obtain  an upper triangular matrix $\bar{\bU}$ whose diagonal 
elements are all equal to $\bigl(\prod_{i=1}^M\lambda_i\bigr)^{1/(2M)}$. 
Finally, we obtain $\bU$ using
\begin{math}
 \bU=\bigl(\prod_{i=1}^M\lambda_i\bigr)^{-1/(2M)}\bar{\bU}. 
\end{math}
Thus, we have established the following proposition:
\begin{proposition}
\label{prop:mse_zf}
The (arithmetic) mean-square error $\tr(\bR_{ee})/M$ of a block-by-block 
transceiver with a ZF-BDFD achieves its minimized lower bound of 
$(M/p_0)\bigl(\prod_{i=1}^M\lambda_i\bigr)^{-1/M}$
when the precoder $\bF=\sqrt{\frac{p_{0}}{M}}\:\tilde{\bV}_M\bPsi_{\zf}$, 
where $\bPsi_{\zf}$ is obtained by applying the algorithm in Appendix~\ref{sec:app_JKZ} 
to $\tilde{\bLambda}_{M}^{1/2}$.
The corresponding feedback matrix $\bB=\bU-\bI$,
 where $\bU$ is the unit-diagonal upper-triangular matrix 
$\bU=\bigl(\prod_{i=1}^M \lambda_i\bigr)^{-1/(2M)}\bar{\bU}$, and
$\bar{\bU}$ is obtained from the QR~decomposition in \eqref{QR3}.
Substituting such $\bF$ and $\bB$ into 
(\ref{equzf1}) yields the feedforward matrix $\bW$.~\hfill$\Box$
\end{proposition}

From the above derivation it is apparent that the precoder in
Proposition~\ref{prop:mse_zf}, which minimizes the arithmetic MSE,
also minimizes the geometric MSE. However, 
a precoder that minimizes the geometric MSE does
not necessarily minimize the arithmetic MSE. 

\subsection{MMSE-BDFD}
\label{sec:deriv_MMSE}
In this subsection, we consider joint transmitter-receiver design for
a system based on the  MMSE-BDFD. The approach is similar to that 
for the ZF-BDFD in  the previous subsection, 
but the details are substantially different.

Recall from Section~\ref{sec:modelling} and Fig.~\ref{fig:model} that 
the received vector is  $\by=\bH\bF\bs+\bv$. 
Hence, the error between $\hat{\bs}$ and $\bs$ is $\be=\bW\by-(\bB+\bI)\bs$. 
The covariance matrix of $\by$ is
\begin{math}
{\bf R}_{yy}=({\bf{HF}})({\bf{HF}})^{H}+{\bf R}_{vv},   
\end{math}
and cross-correlation matrix of $\bs$ and $\by$ is
\begin{math}
  {\bf R}_{sy}=({\bf{HF}})^{H}={\bf R}_{ys}^{H}.  
\end{math}
In order to determine the minimum MSE feedforward matrix, $\bW_{\mmse}$, 
we exploit the standard first-order necessary condition for optimality known as the
orthogonality principle~\cite{proakis}, namely
\begin{math}
   E[\be\by^H]
              =  {\bf W}\bR_{yy}-({\bf B}+{\bf I})\bR_{sy}=\boldsymbol{0}.
\end{math}
Therefore, 
\begin{eqnarray}\label{equww}
 {\bf W}_{\mmse} & = & ({\bf B}+{\bf I}){\bf R}_{sy}{\bf R}_{yy}^{-1}.
\end{eqnarray} 
Substituting (\ref{equww}) into (\ref{Reet}), and invoking the
Matrix Inversion Lemma 
$(\bA+\bC\bB^{-1}\bD)^{-1}=
\bA^{-1}-\bA^{-1}\bC(\bB+\bD\bA^{-1}\bC)^{-1}\bD\bA^{-1}$, \cite{magnus}, 
the covariance matrix of the error can be written as
    \begin{equation}\label{REE}
     {\bf R}_{ee,\mmse}=({\bf B}+{\bf I})\left(\bI+\bF^H\bH^H\bR_{vv}^{-1}\bH\bF\right)^{-1}({\bf B}+{\bf I})^H.
    \end{equation}
Our goal is to design the $\bF$ and $\bB$ to minimize the 
MSE subject to the power constraint. 
Letting $\bU=\bB+\bI$, the design problem (\ref{opt1}) can be rewritten as 
     \begin{subequations}
\begin{align}
\min_{{\bf F},{\bU}} & \quad {\operatorname{tr}}\left({\bf U}\left(\bI+\bF^H\bH^H\bR_{vv}^{-1}\bH\bF\right)^{-1}{\bf U}^H\right)\label{ORGP1}  \\
\mbox{subject to} & \quad   {\operatorname{tr}}({\bf FF}^{H}) \leq p_0, 
\;\text{and}\;\\
&\quad
\bU\;\text{being a unit-diagonal upper-triangular matrix.}    
\end{align}
\end{subequations}

Following the first stage outlined at the beginning
of  Section~\ref{sec:deriv},
we now obtain and minimize a lower bound on the MSE. 
According to the trace-determinant inequality \eqref{eq:AGMI},
we have that
  \begin{align} 
      \!&{\operatorname{tr}}\left({\bf U}\left(\bI+\bF^H\bH^H\bR_{vv}^{-1}\bH\bF\right)^{-1}{\bf U}^H\right) \notag\\
& \qquad\qquad\geq  M 
\bigl|\bU\left(\bI+\bF^H\bH^H\bR_{vv}^{-1}\bH\bF\right)^{-1}\bU^H\bigr|^{1/M}\notag\\
       &\qquad\qquad =  M \left|\bI+\bF^H\bH^H\bR_{vv}^{-1}\bf{HF}\right|^{-1/M}.
\label{eq:geo_MSE_MMSE}
\end{align}    
Therefore, the lower bound on the MSE can be minimized by solving:
\begin{subequations}\label{LOW1}
\begin{align}
\max_{{\bf F}} & \quad |\bI+\mathbf{F}^H\mathbf{H}^H\bR_{vv}^{-1}\mathbf{H}\mathbf{F}|\\
\mbox{subject to} & \quad   {\operatorname{tr}}({\bf FF}^{H}) \leq p_0.    
\end{align}
\end{subequations}
As in the ZF case, the problem of minimizing the lower bound depends only
on the transmitter. 
We point out that the objective in (\ref{LOW1}a) is 
equivalent to minimizing the geometric MSE implicit 
in \eqref{eq:geo_MSE_MMSE}.
Furthermore, the logarithm of the objective in (\ref{LOW1}a) is the 
 mutual information between the transmitter and receiver for
Gaussian signals. (An analogous observation has been made in several
similar contexts~\cite{salz,falconer-foschini,cioffi,Yang,cioffi-forney-GDFE}.)
Hence, minimizing the lower bound on the arithmetic MSE in \eqref{eq:geo_MSE_MMSE} 
is equivalent to maximizing the Gaussian mutual information.

Given that the problem in \eqref{LOW1} is equivalent to
maximizing the mutual information for Gaussian signals, the
solution involves a ``waterfilling'' power allocation over
the eigenvectors of $\mathbf{H}^H\mathbf{R}_{vv}^{-1}\mathbf{H}$,~\cite{witsenhausen}.
More formally, the solution depends on a parameter 
$r\leq K$  which is the largest integer satisfying
\begin{math}\label{FAI2}
    {1}/{\lambda_{r}} < \bigl(p_{0}+\sum_{j=1}^r \lambda_{j}^{-1}\bigr)/r.
  \end{math}
If we define $q=\min\{r,M\}$, then  the following set of precoders%
\footnote{If $M=K$ and $r=K$, or if $\lambda_q>\lambda_{q+1}$, 
this set is the set of all precoders that minimize the lower bound.}
 minimize the lower bound~\cite{witsenhausen},
$\bF=\tilde{\bV}_q\begin{bmatrix}\boldsymbol{\Phi} &
\boldsymbol{0}_{q\times (M-q)} \end{bmatrix}\boldsymbol{\Psi}$, where 
$\boldsymbol{\Phi}$ is a $q \times q$ diagonal matrix with diagonal elements satisfying
\begin{equation}\label{MMSEFAI}
   |\phi_{ii}|^2=
 \frac{1}{q}\Bigl(p_{0}+\sum_{j=1}^{q} \lambda_{j}^{-1}\Bigr)-\lambda_{i}^{-1}, 
\end{equation}
 and $\boldsymbol{\Psi}$ is an arbitrary $M\times M$ unitary matrix.%
 \footnote{The rank of the  resulting product $\mathbf{H}\bF$ 
is $q$, and hence if $M$ were a design variable rather than
a parameter of the problem, a natural choice for $M$ would be  $M=r$.}
 In that case, the minimal value of the
lower bound on the MSE generated by  \eqref{eq:geo_MSE_MMSE} 
and \eqref{LOW1} is 
\begin{equation}
\label{MMSELB}
      \bar{e}_{\mmse}^{2} \geq q^{q/M}
\Bigl(p_{0}+\sum_{j=1}^{q} \lambda_j^{-1}\Bigr)^{-q/M}
\prod_{j=1}^{q} \lambda_j^{-1/M},
   \end{equation}
which is independent of our design parameters $\bF$ and $\bB$.

Moving to the second stage of our general approach, we now determine
a transceiver that achieves the 
minimized lower bound in \eqref{MMSELB}.
For ease of exposition, we define
$\breve{\bPhi}=\left[\begin{smallmatrix}\boldsymbol{\Phi} &
\boldsymbol{0}_{q\times (M-q)}\end{smallmatrix}\right]$.
Substituting $\bF=\tilde{\bV}_q\breve{\bPhi}\boldsymbol{\Psi}$ 
into (\ref{REE}) and (\ref{ORGP1}), the arithmetic MSE is $\tr(\bR_{ee,\mmse})/M$, where  
\begin{equation}\label{REE1}
  \bR_{ee,\mmse}={\bf U}\bPsi^H(\bI_M+\breve{\bPhi}^T\tilde{\boldsymbol{\Lambda}}_{q}
  \breve{\bPhi})^{-1}\bPsi{\bf U}^H.
\end{equation}
Using the trace-determinant inequality \eqref{eq:AGMI}, 
for the MSE to achieve its minimized lower bound,
we must choose $\bU$ and $\boldsymbol{\Psi}$ so that 
$\bR_{ee,\mmse}=\check{\sigma}_e^2\bI$, where 
\begin{math}
  \check{\sigma}_e^2=
   q^{q/M}\bigl(p_{0}+\sum_{j=1}^{q}\lambda_j^{-1}\bigr)^{-q/M} 
  \prod_{j=1}^{q}
   \lambda_j^{-1/M}.
\end{math}
That is, a system of the form in \eqref{MMSEFAI} achieves the
minimized lower bound on the MSE in \eqref{REE1} if and only if we can find
$\check{\bU}=(1/\check{\sigma}_e)\bU$ 
and unitary matrices $\boldsymbol{\Psi}$ and $\bQ$ so that
 \begin{equation}\label{MMSEQRA}
   (\bI_M+
   \breve{\bPhi}^T\tilde{\boldsymbol{\Lambda}}_{q}
   \breve{\boldsymbol{\Phi}})^{1/2}
\bPsi=\bQ\check{\bU}.
\end{equation}
According to  Lemma~\ref{lem:JKZ}, 
there exists a unitary matrix $\bPsi$
 such that the QR~decomposition of 
$(\bI_M+\breve{\bPhi}^T\tilde{\boldsymbol{\Lambda}}_{q}
   \breve{\boldsymbol{\Phi}})^{1/2} \bPsi$
has an upper triangular ``R-factor'' with diagonal elements all equal to 
$|(\bI_M+   \breve{\bPhi}^T\tilde{\boldsymbol{\Lambda}}_{q}
   \breve{\boldsymbol{\Phi}})^{1/2}\bPsi|^{1/(2M)}$. 
This unitary matrix can be obtained by applying the 
algorithm in Appendix~\ref{sec:app_JKZ} to 
$(\bI_M+\breve{\bPhi}^T\tilde{\boldsymbol{\Lambda}}_{q}
   \breve{\boldsymbol{\Phi}})^{1/2}$.
We summarize this  result in the following proposition.
\begin{proposition} 
\label{prop:mse_mmse}
The mean-square error $\tr(\bR_{ee})/M$ for a block-by-block 
transceiver with an MMSE-BDFD achieves its minimized lower 
bound (\ref{MMSELB}) when the precoder 
$\bF=\tilde{\bV}_{q}
\left[\begin{smallmatrix}\boldsymbol{\Phi} & \boldsymbol{0}_{q\times (M-q)}
\end{smallmatrix}\right]\bPsi_{\mmse}$, 
where $\boldsymbol{\Phi}$ satisfies (\ref{MMSEFAI}),
and $\bPsi_{\mmse}$ is  obtained by applying the algorithm in Appendix~\ref{sec:app_JKZ}
to $(\bI_M+\breve{\bPhi}^T\tilde{\bLambda}_{q}\breve{\bPhi})^{1/2}$.
The corresponding feedback matrix  $\bB=\bU-\bI$, 
where $\bU$ is the unit-diagonal upper-triangular matrix 
$\bU=\check{\sigma}_e \check{\bU}$ and $\check{\bU}$ is obtained from the
QR~decomposition in \eqref{MMSEQRA}.
Substituting such $\bF$ and $\bB$ into (\ref{equww}) 
yields the feedforward matrix $\bW$.   \hfill$\Box$
\end{proposition} 

As was the case for the ZF-BDFD in Section~\ref{sec:deriv_ZF}, the
precoder in Proposition~\ref{prop:mse_mmse}, which minimizes the arithmetic MSE, 
lies within the set of precoders that minimize the geometric MSE, but a
precoder chosen arbitrarily from the set of precoders that minimize the
geometric MSE does not necessarily minimize the arithmetic
MSE. This observation provides a connection between
the proposed design and an earlier design for a more
general overlapping block transmission system
in which the transmitter was designed to minimize
the geometric MSE~\cite{Yang}.
In the context of the block-by-block transmission schemes 
that we have considered, the design in~\cite{Yang}
 corresponds to choosing $\bPsi=\bI_M$, 
rather than choice of $\bPsi=\bPsi_{\mmse}$ in
Proposition~\ref{prop:mse_mmse}.
While the choice of $\bPsi=\bI_M$ results in a system that
minimizes the geometric MSE, it does not minimize the arithmetic
MSE in the general case. In addition, the SINR for each element of the
block may be different. In contrast, the choice of $\bPsi=\bPsi_{\mmse}$
minimizes the geometric MSE and the arithmetic
MSE, and provides an equal SINR for each element of the block.

The choice of $\bPsi$ also has an impact on the nature of coding strategies
for approaching the capacity of the block-by-block transmission system.
From the discussion following \eqref{LOW1} it is evident that the
Gaussian mutual information is maximized by choosing $M=r$
and  employing a transmitter matrix of
the form $\bF = \tilde{\bV}_r \bPhi \bPsi$, where 
$\bPhi$ satisfies \eqref{MMSEFAI} and $\bPsi$ is
an arbitrary $r\times r$ unitary matrix.
Since the MMSE-BDFD is a ``canonical'' receiver%
\footnote{The term  ``canonical'' is used to denote the fact that in the absence 
of error propagation, employing an MMSE-BDFD in place of the optimal
detector does not reduce the achievable data rate~\cite{cioffi,cioffi-forney-GDFE}. 
Methods for exploiting this property of the MMSE-BDFD were
 described in~\cite{Varanasi_Guess_Asilomar97,Guess_Varanasi_ISIT2000}.}
 for Gaussian signals~\cite{cioffi,cioffi-forney-GDFE,Guess_Varanasi_MMSE_DFE},
this suggests that by using sufficiently powerful codes,
reliable communication at rates approaching the capacity of the
block transmission system can be achieved by employing any $\bF$ of
this form and the MMSE-BDFD~\cite{cioffi,cioffi-forney-GDFE,Guess_Varanasi_MMSE_DFE}.
The choice $\bPsi=\bI_r$  results in
a ``vector coding'' scheme~\cite{cioffi-forney-GDFE,Kasturia,Lechleider,scaglione1,palomar}
in which the feedback component of the MMSE-BDFD is inactive; i.e., $\bB=\bZero$.
Vector coding induces an equivalent system with $r$ parallel Gaussian subchannels,
each with a possibly different SNR~$\rho_i$. 
(Standard discrete multitone  (DMT) modulation schemes~\cite{chow,bingham}
are a class of vector coding schemes.) Therefore, 
one can approach the capacity of the block transmission scheme by choosing
the  code for the $i$th element of the block to be one that  approximates the 
ideal Gaussian code of rate $b_i=\log_2(1+\rho_i)$ bits per channel use.
(Such approximations will often involve 
the selection of a constellation for each element of the block.)
The choice $\bPsi=\bPsi_{\mmse}$ results in a system in which 
the feedback component of the MMSE-BDFD is active, and
the inputs to the decision device are uncorrelated and have identical SINRs~$\rho$.
Since the MMSE-BDFD is a canonical receiver, 
this suggests that one can also approach the capacity
of the block transmission system by employing 
an independent instance of the same approximation of the 
ideal Gaussian code of rate $b=\log_2(1+\rho)$ for each element of
the block. The MMSE-BDFD used when $\bPsi=\bPsi_{\mmse}$ 
is more complicated to implement than the linear
detector of the vector coding scheme because of the need to compute
the feedback signal. However, the vector coding approach requires
the design  (and implementation) of (up to) $r$ codes, 
one for each element of the block, whereas  the proposed  design
requires the design of only one code.

\section{Bit Error Rate Performance}
\label{sec:BER}
In this section, we show that the $(\bF,\bB)$ pairs designed in
Section~\ref{sec:deriv} to minimize the arithmetic MSE also 
minimize the (dominant components of the uncoded) bit error rate (BER)
of a block transmission system with uniform bit loading at moderate-to-high
block SNRs. We define the  average BER of the detected signal to be
 the average of the probability of error of each element of the block; i.e.,
\begin{equation}\label{Pe}
   P_e={\frac{1}{M}}\sum_{i=1}^MP_{e,i},
\end{equation}
where $ P_{e,i}$ denotes the BER of the $i$th symbol $s_{i}$.
For ease of exposition, we will deal with the ZF and MMSE-BDFDs
separately. We will begin with the case
of the ZF-BDFD.

\subsection{ZF-BDFD}
\label{sec:BER_ZF}

For the ZF-BDFD%
\footnote{We implicitly assume that
$\operatorname{rank}(\bH)\geq M$ so that the ZF-BDFD exists.}
 and for square%
\footnote{For notational simplicity we have restricted our attention
to square QAM constellations. The extension to rectangular QAM
constellations can be derived in a straightforward
manner using the BER expressions in \cite{Cho,Cho_EL}.}
QAM signalling  with $2b_i$ bits per symbol, if all the previous decisions 
are correct $P_{e,i}$ is closely approximated%
\footnote{In the case of QPSK signalling,  the expression in \eqref{Pei}, 
in which $\zeta_i=0$, is exact.}
by~\cite{Cho}
\begin{align}
P_{e,i} \approx \tilde{P}_{e,i} &=
 \alpha_i \operatorname{erfc} \big(
\sqrt{\beta_i\rho_{i,\zf}}\big)
+ \zeta _i  \operatorname{erfc}
\big( 3 \sqrt{\beta_i\rho_{i,\zf}}\big), \label{Pei}
\end{align}
where $\operatorname{erfc}(x)=(2/\sqrt{\pi})\int_x^{\infty}
e^{-z^2}\,dz$ is the error function complement,
$\rho_{i,\zf}$ is the decision point SNR
for the $i$th symbol in the block, 
$\alpha_i =  \frac{\sqrt{4^{b_i}}-1}{b_i\sqrt{4^{b_i}}}$,
$\beta_i= \frac{3b_i}{(4^{b_i}-1)}$,
and $\zeta_i= \frac{\sqrt{4^{b_i}}-2}{b_i\sqrt{4^{b_i}}}$.
Hence,
\begin{equation*}
P_e\approx \tilde{P}_e = \frac{1}{M} \sum_{i=1}^M \tilde{P}_{e,i}.
\end{equation*}
Under the assumption that all the previous symbols were correctly detected, we have that
 \begin{equation} \label{eq:rhoi_zf1}
\rho_{i,\zf}= \frac{E[s_{i}^2]}{E[| {\hat s_{i}}-s_{i}|^2]},
 \end{equation}
and under our assumptions 
that $E[\bs\bs^H]=\bI$ and $E[\bs\bv^H]=\mathbf{0}$, 
this expression simplifies to
\begin{equation}\label{SNRi}
        \rho_{i,\zf}  =  \frac{1}{[\bR_{ee,\zf}]_{ii}}.
\end{equation}
Therefore, the average BER can be closely approximated by 
\begin{multline}\label{Peo}
   P_e\approx\tilde{P}_e=
\frac{1}{M}\sum_{i=1}^M
\alpha_i\operatorname{erfc}\left(\sqrt{{\beta_i}/{[\bR_{ee,\zf}]_{ii}}}\right)
\\
+\zeta_i\operatorname{erfc}\left(3\sqrt{{\beta_i}/{[\bR_{ee,\zf}]_{ii}}}\right).
\end{multline}
Since our precoders generate equal decision point SNRs for each element of the
block, we will assume uniform bit-loading in the remainder of this section,
and therefore we will drop the element index, $i$, in $\alpha_i$, $\beta_i$
and $\zeta_i$. When $[{\bf R}_{ee,\zf}]_{ii}<2\beta/3$,
which corresponds to moderate-to-high SNRs, $\tilde{P}_{e}$ is a convex 
function of $[{\bf{R}}_{ee}]_{ii}$, \cite{dingy,yanwu_globecom,palomar}. 
By applying Jensen's inequality~\cite{cover} to (\ref{Peo}), 
we obtain the following lower bound on the average BER
    \begin{multline}
       \tilde{P}_e  \geq 
\alpha\, \erfc\Bigl(\sqrt{\beta M/\operatorname{tr}(\bR_{ee,\zf})}\Bigr)
\\
+ 
\zeta \,\erfc\Bigl(3\sqrt{\beta M/\operatorname{tr}(\bR_{ee,\zf})}\Bigr).
\label{Pelb}
    \end{multline}
Equality in (\ref{Pelb}) holds if and only if the diagonal elements of 
${\bf R}_{ee,\zf}$ are equal. 

Equation ({\ref{Pelb})  exposes an intriguing relationship 
between the (arithmetic) MSE and the BER. 
Since minimizing $\operatorname{tr}(\bR_{ee,\zf})$ simultaneously
minimizes both terms in the summation 
on the right hand side of (\ref{Pelb}), 
minimizing the lower bound on $\tilde{P}_e$
in (\ref{Pelb}) is equivalent to minimizing the MSE;
i.e., it is equivalent to minimizing $\operatorname{tr}(\bR_{ee,\zf})$. 
Therefore, the lower bound on $\tilde{P}_e$ achieves its minimum value if the MSE is minimal. 
However, for the actual $\tilde{P}_e$ to achieve its lower bound 
(i.e., for (\ref{Pelb}) to hold with equality), 
the diagonal elements of $\bR_{ee,\zf}$ must be identical.%
\footnote{The alternative analysis in~\cite{varanasi} generates a
related observation.} 
Fortunately, the design proposed in Proposition~\ref{prop:mse_zf} results in
$\bR_{ee,\zf}=\sigma_e^2\bI$, and hence the proposed design,
which minimizes the (arithmetic) MSE of a ZF-BDFD, 
also minimizes the BER of the ZF-BDFD at moderate-to-high SNRs, 
in the sense that it minimizes $\tilde{P}_e$ in \eqref{Peo}.

\subsection{MMSE-BDFD}
\label{sec:BER_MMSE}
The analysis of the previous section can be extended to the case of the MMSE-BDFD
if the residual intra-block interference on each element of the
block is approximated by  a Gaussian random variable. For large block sizes,
this approximation is (almost surely) sufficiently accurate 
for all but the last few elements of the block
(c.f.,~\cite{PoorVerdu,ZhangChongTse,Guo_Ras_Verdu}), and hence it is
appropriate for our analysis.
In order to account for the bias in the MMSE-BDFD
(e.g.,~\cite{cioffi}), we can express the BER as a function of the 
decision point SINR of the $i$th element of
the block~\cite{cioffi,cioffi-forney-GDFE,palomar},
\begin{equation}
\rho_{i,\mmse} = \frac{1}{[\bR_{ee,\mmse}]_{ii}}-1.
\end{equation} 
(Note that $0\leq [\bR_{\mmse}]_{ii}<1$.)
By replacing $\rho_{i,\zf}$ in \eqref{Pei} by $\rho_{i,\mmse}$,
the BER of the MMSE-BDFE can  be approximated by 
\begin{multline}\label{Peo_mmse}
   P_e\approx\tilde{P}_e=
\frac{1}{M}\sum_{i=1}^M
\alpha_i\operatorname{erfc}\Bigl(\sqrt{{\beta_i}\bigl(([\bR_{ee,\mmse}]_{ii})^{-1}-1\bigr)}\Bigr)
\\
+\zeta_i\operatorname{erfc}\Bigl(3\sqrt{{\beta_i}\bigl(([\bR_{ee,\mmse}]_{ii})^{-1}-1\bigr)}\Bigr).
\end{multline}
As was the case for the ZF-BDFD, this function is convex 
in $[\bR_{ee,\mmse}]_{ii}$ when $[\bR_{ee,\mmse}]_{ii}$ is
below a (reasonably large) threshold~\cite{chan,palomar}, and hence 
for a system in which uniform bit loading is applied, Jensen's inequality
can be used to show that 
\begin{multline}
\tilde{P}_e \geq
\alpha\operatorname{erfc}\Bigl(\sqrt{\beta\bigl(M/\operatorname{tr}(\bR_{ee,\mmse})-1\bigr)}\Bigl)
\\
+ \zeta\operatorname{erfc}\Bigl(3\sqrt{\beta\bigl(M/\operatorname{tr}(\bR_{ee,\mmse})-1\bigr)}\Bigl),
\label{eq:BER_lowerbound_MMSE}
\end{multline}%
with equality holding when the diagonal elements of $\bR_{ee,\mmse}$ are equal.
Hence, using similar arguments to those used in the case of the ZF-BDFD,
the design proposed in Proposition~\ref{prop:mse_mmse},
which minimizes the arithmetic MSE of the MMSE-BDFD and results in 
$\bR_{ee,\mmse} = \check{\sigma}_e^2\bI$, 
also minimizes the BER of the MMSE-BDFD at moderate-to-high SNRs, 
in the sense that it minimizes $\tilde{P}_e$ in \eqref{Peo_mmse}.%
\footnote{Note that if $M>r$, then $\operatorname{rank}(\bF)<M$
and hence the lower bound on the BER in \eqref{eq:BER_lowerbound_MMSE} will be quite high.
If $M$ were a design variable, rather than a parameter of the problem, reducing the
symbol rate to $M=r$ would result in a substantial reduction
in the error rate of the optimized system.}

\section{Performance Analysis}
\label{sec:sims}

In Section~\ref{sec:BER} it was shown that the precoders that we
designed in Section~\ref{sec:deriv} (essentially) minimize the BER of the BDFD,
under the assumption that the decisions that are fed back in the receiver 
are  correct. It can also be shown (see Appendix~\ref{sec:app_BERcomp})
that under the same assumption
the optimized system for an MMSE-BDFD provides a lower BER than
the optimized system for a ZF-BDFD, and that each optimized BDFD system
provides a lower BER than the optimized system for the corresponding linear
detector; c.f.,~\cite{chan,dingy,palomar}.
That said, an incorrect decision in a BDFD
can make it more likely that subsequent errors will 
 occur by feeding back incorrect decisions. This may lead to error propagation
across the block. (Recall that error propagation between blocks is
explicitly avoided in block-by-block communication systems.)
A standard bound on the probability of error of a conventional
decision feedback equalizer in the presence of error
propagation is a simple  multiple of the probability of error 
in the absence of error propagation~\cite{dutt}. 
This suggests that the systems designed in Section~\ref{sec:deriv}
should perform well in the presence of error propagation.
(A bound that is sometimes tighter~\cite{Beaulieu} 
generates similar insight.)
 In this section, we seek to verify these suggestions by analyzing, 
 via simulation, the (uncoded) BER performance of the system when 
 error propagation may occur.

We will consider two communication scenarios: zero-padded
block transmission~\cite{scaglione,scaglione1,stamoulis}
 through a (quasi-static) scalar finite impulse response (FIR)
frequency-selective fading channel that is constant over the
length of the block; and 
transmission through a narrowband (i.e., frequency-flat) multiple
antenna fading channel with at least as many receive antennas as
transmit antennas~\cite{foschini}. 
In the first scenario, the channel matrix $\bH$ is a tall,
lower triangular, Toeplitz matrix, but in the second
scenario $\bH$ does not possess any deterministic structure.
We will evaluate the average BER
performance of various transceivers for these channels
in the presence of additive white Gaussian noise 
at the receiver; i.e., $\bR_{vv}=\sigma^2\bI$.
We will plot the BER performance curves as a function of
the (system) SNR, which we define as being the ratio of the
transmitted energy per symbol to the noise variance; 
i.e., $(p_0/M)/\sigma^2$.

In addition to the transceivers we designed for the ZF-BDFD and MMSE-BDFD
in Section~\ref{sec:deriv}, for which the precoders are denoted by 
$\bF_{\text{\textsc{opt-zf-bdfd}}}$ and 
$\bF_{\text{\textsc{opt-mmse-bdfd}}}$, respectively,
when $M=K$  we will also consider the direct transmission scheme,
for which the precoder is
\begin{equation}
\bF_{\text{\textsc{i}}}=\sqrt{{p_0}/{M}}\:\bI_{M},
\label{eq:FI}
\end{equation}
 and the discrete Fourier transform (DFT) precoded scheme,
for which the precoder is
\begin{equation}
\bF_{\text{\textsc{dft}}}=\sqrt{{p_0}/{M}}\:\bD^H,
\label{eq:FDFT}
\end{equation}
where $\bD$ is the  normalized $M\times M$ DFT matrix.
For the precoders in \eqref{eq:FI} and \eqref{eq:FDFT},
 the receiver matrices $\bB$ and $\bW$ are chosen according
 to the (separate) design procedures 
for the ZF-BDFD and MMSE-BDFD in \cite{stamoulis}. 
(Note that the precoders in the direct and DFT schemes
are channel independent.)
For all these precoders, we provide BER curves 
for the idealized detector, in which the decisions that are
fed back are correct, and for the practical detector, in which 
the actual decisions are fed back (and hence error
propagation may occur).

In order to assess the extent of the performance gains 
(derived in Appendix~\ref{sec:app_BERcomp}) of the optimized BDFD systems
over the optimized system for the corresponding linear detector,
we will include the performance 
of systems with linear ZF and MMSE detection and precoders
designed so that the 
BER at moderate-to-high block SNRs is minimized~\cite{dingy,chan,palomar}.
Using the notational conventions in Sections~\ref{sec:modelling} and \ref{sec:deriv},
in particular the ordered eigen decomposition 
$\bH^H\bR_{vv}^{-1}\bH=\bV\boldsymbol{\Lambda}\bV^H$,
a minimum BER precoder for the linear ZF detector is~\cite{dingy}
\begin{align}
\bF_{\text{\textsc{opt-zf-l}}} &= \sqrt{p_0/\operatorname{tr}(\tilde{\bLambda}_M^{-1/2})}\; 
\tilde{\bV}_M \tilde{\bLambda}_M^{-1/4} \bD,\\
\intertext{and  one for the linear MMSE detector is~\cite{chan,palomar}}
\bF_{\text{\textsc{opt-mmse-l}}} &= \tilde{\bV}_k 
\begin{bmatrix}\boldsymbol{\Upsilon} & \bZero_{k\times(M-k)}\end{bmatrix}\bD,
\label{eq:Fopt_linMMSE}
\end{align}
where the integer $k = \min\{\ell,M\}$, where $\ell$ is
the largest integer such that
\begin{equation*}
\lambda_{\ell}^{-1/2} \Bigl(\sum_{j=1}^{\ell} \lambda_j^{-1/2}\Bigr)
- \sum_{j=1}^{\ell} \lambda_j^{-1} < p_0,
\end{equation*}
and $\boldsymbol{\Upsilon}$ is a $k\times k$
diagonal matrix with diagonal elements satisfying 
\begin{equation*}
|\upsilon_{ii}|^2 = 
\biggl(\frac{p_0 + \sum_{j=1}^{k} \lambda_j^{-1}}%
{\sum_{j=1}^{k} \lambda_j^{-1/2}}\biggr)\lambda_i^{-1/2} - \lambda_i^{-1}.
\end{equation*}

\subsection{Scalar frequency-selective fading channel}
\label{sec:FIR}

In this section we consider the case of zero-padded block transmission through a
(quasi-static) scalar FIR frequency-selective fading channel.
In this case,
the direct transmission scheme in \eqref{eq:FI} is sometimes referred
to as the ``single-carrier zero-padded'' (SCZP) scheme~\cite{wang3},
and the DFT precoded scheme is sometimes called the ``zero-padded OFDM'' (ZP-OFDM) 
scheme~\cite{muquet}.
We consider a scenario in which the channel is of length $L+1=5$
and $L$ zeros are appended to each block of channel symbols~$\bu$. 
The symbol block $\bs$ is of length $M=16$, and we consider square
precoders~$\bF$. (Hence, $K=16$ and $P=K+L=20$.)
Each element of $\bs$ is an independently selected symbol from the
4--QAM constellation, with each constellation point being
equally likely. In Fig.~\ref{fig:BER_ZFBDFD_random} we plot the  BER
for the ZF-BDFD transceivers, averaged over ten thousand
channel realizations. (In the optimized designs, the transceiver 
was re-designed for each channel realization.)
For each channel realization  the tap coefficients were generated
independently from a zero-mean circular complex Gaussian distribution
and then normalized so that
the impulse response had unit energy.
It is clear from the solid curves in Fig.~\ref{fig:BER_ZFBDFD_random}
 that in the absence of error propagation,
the design proposed in Proposition~\ref{prop:mse_zf}
performs better than all the other transmission schemes,%
\footnote{As predicted by the derivation in
Section~\ref{sec:BER_ZF}, the proposed precoder performs
better than all other transmission schemes for each
realization of the channel.}
although the SNR gain over the direct transmission (SCZP) scheme is rather small
(around 0.5~dB at a BER of $10^{-4}$).
Furthermore, the dashed curves demonstrate that this performance advantage
is maintained in the presence of error propagation.
In particular, the performance of the proposed scheme in
the presence of error propagation is as good as the performance
of the SCZP scheme in the absence of error propagation.
The combination of the DFT transmitter (ZP-OFDM) and the ZF-BDFD performs poorly at
moderate-to-high block SNRs. In fact, it is apparent from
Fig.~\ref{fig:BER_ZFBDFD_random} that   the linear ZF detection scheme with its
minimum BER precoder~\cite{dingy}  performs better than the combination of the
DFT transmitter and the ZF-BDFD. 
However, as predicted by the analysis in Appendix~\ref{sec:app_BERcomp},
the optimal precoder for the ZF-BDFD provides substantially better performance
than the combination of the linear ZF detector and its minimum BER precoder.

The corresponding results for the MMSE-BDFD are provided in
Fig.~\ref{fig:BER_MMSEBDFD_random}.
The same trends are observed and the SNR gains are 
at least as large. Furthermore, the improved BER performance of the
optimized MMSE-BDFD system over the optimized ZF-BDFD system
predicted by the analysis in Appendix~\ref{sec:app_BERcomp}
can be clearly observed.
In both Figs~\ref{fig:BER_ZFBDFD_random} and \ref{fig:BER_MMSEBDFD_random},
the performance of the optimized scheme in the absence of error
propagation is indistinguishable from the corresponding 
bound on $\tilde{P}_e$ in Section~\ref{sec:BER}; 
c.f.,~\eqref{Pelb} and \eqref{eq:BER_lowerbound_MMSE}, respectively.

\begin{figure}
   \centering
      \resizebox{0.48\textwidth}{!}{\includegraphics{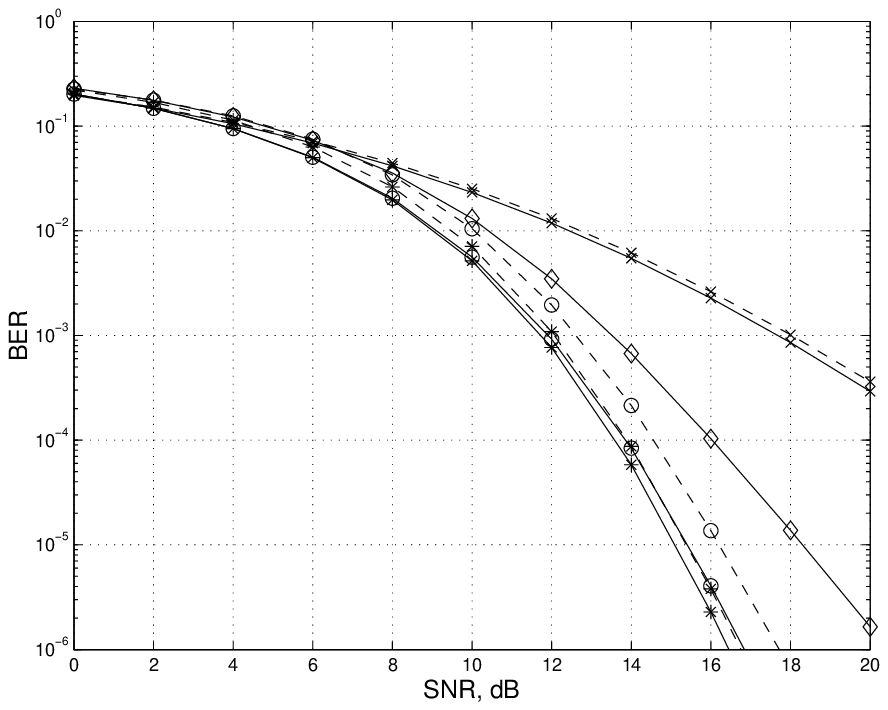}}
      \caption{Average BER performance of the ZF-BDFD for
the various precoders  and the linear ZF detector with its optimal
precoder in the scalar frequency-selective fading channel scenario
 in Section~\ref{sec:FIR}. The solid curves denote
performance achieved in the absence of error propagation, and the
dashed curves incorporate the effects of error propagation. 
Legend---$\ast$: optimized scheme,
$\bF_{\text{\textsc{opt-zf-bdfd}}}$;
$\circ$: direct (SCZP), $\bF_{\text{\textsc{i}}}$; 
$\times$: DFT (ZP-OFDM), 
$\bF_{\text{\textsc{dft}}}$;
$\diamond$: optimized linear ZF scheme, 
$\bF_{\text{\textsc{opt-zf-l}}}$.}
\label{fig:BER_ZFBDFD_random}
\end{figure}

\begin{figure}
   \centering
      \resizebox{0.48\textwidth}{!}{\includegraphics{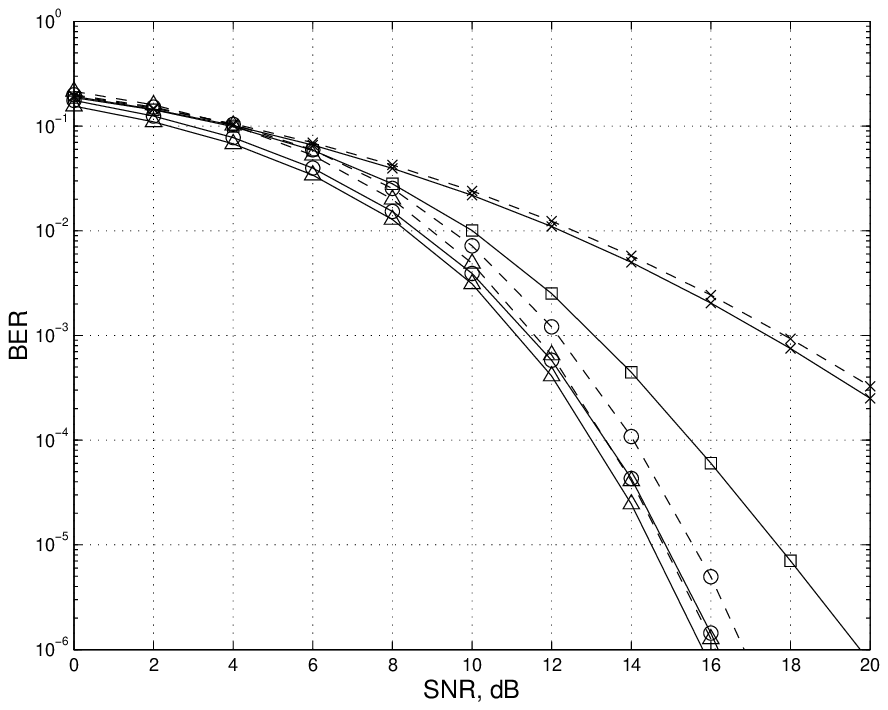}}
      \caption{Average BER performance of the   MMSE-BDFD for
the various precoders
and the linear MMSE detector with its optimal precoder
in the scalar frequency-selective fading channel scenario
 in Section~\ref{sec:FIR}. The solid curves denote
performance achieved in the absence of error propagation, and the
dashed curves incorporate the effects of error propagation. 
Legend---$\vartriangle$: optimized scheme,
$\bF_{\text{\textsc{opt-mmse-bdfd}}}$;
$\circ$: direct (SCZP), $\bF_{\text{\textsc{i}}}$; 
$\times$: DFT (ZP-OFDM), $\bF_{\text{\textsc{dft}}}$;
$\Box$: optimized linear MMSE scheme, $\bF_{\text{\textsc{opt-mmse-l}}}$.}
\label{fig:BER_MMSEBDFD_random}
\end{figure}

An interesting by-product of the above performance evaluation
is the good performance provided by the 
(channel independent) direct transmission scheme (SCZP).
In fact, the SCZP scheme is an optimal channel independent
transmission scheme for systems 
that employ linear~\cite{Lin-Phoong} or
maximum likelihood~\cite{wang3,JKZ_opt_precoder} detection,
and it approaches the diversity-multiplexing 
trade-off for a standard class of FIR channels
as the block length grows~\cite{GrokopTse}.
These desirable characteristics are due, in part, to the fact that the SCZP
scheme preserves the good conditioning properties implicit in
the tall lower-triangular Toeplitz structure of the channel matrix.

\subsection{Multiple antenna systems}
\label{sec:MIMO}

In this example, we consider the case of narrowband transmission over a multiple
antenna channel with at least as many receiver antennas as transmitter antennas. In
this scenario, the combination of the direct transmission scheme and a BDFD
is sometimes referred to as (uncoded) V-BLAST with a (fixed-order) 
``nulling and cancelling'' receiver~\cite{biglieri,foschini,ginis}.
We consider
a standard Rayleigh model for the channel in which the paths between antennas
are modelled as independent zero-mean circular Gaussian random  variables of
unit variance. 

We will focus on scenarios with $K=3$ transmitter 
antennas and $P=3$ or $4$ receiver antennas in which
$M=K=3$ symbols are transmitted per channel use.
Each element of $\bs$ is an independent and equally-likely 4--QAM symbol.
Therefore, the bit rate of each scheme is 6 bits-per-channel-use (bpcu).
In Figs~\ref{fig:ZF_MIMO_M3P3} and \ref{fig:ZF_MIMO_M3P4}, 
we plot the average BER performance over ten thousand channel realizations
 of the various transmission schemes with the ZF receivers, 
 and in Figs~\ref{fig:MMSE_MIMO_M3P3} and \ref{fig:MMSE_MIMO_M3P4}
we plot the corresponding curves for the MMSE receivers. While most of the basic
trends from the case of the scalar frequency-selective channels are maintained
in the multiple antenna scenario, the performance advantages of the
precoders designed in Section~\ref{sec:deriv} are much greater.
(The SNR gains are of the order of 6--8~dB at a BER of $10^{-4}$.)
This can be attributed to the fact that the channel matrix $\bH$ 
does not possess any deterministic structure. 
In particular, the probability of encountering a channel matrix that does not have
$M$ substantial singular values is not negligible.
Since the proposed designs provide significantly
better performance in those cases, the average
performance is also substantially improved.

As expected, the performance of the optimized ZF-BDFD scheme in the absence
of error propagation in Figs~\ref{fig:ZF_MIMO_M3P3} and \ref{fig:ZF_MIMO_M3P4}
is equal to the lower bound on $\tilde{P}_e$ in \eqref{Pelb}.
(Recall that we are using 4-QAM signalling.) 
However, in the MMSE-BDFD case, the lower bound on $\tilde{P}_e$
in \eqref{eq:BER_lowerbound_MMSE} is distinguishable from the simulated BER 
in the absence of error propagation. This is due to the fact that the block size ($M=3$) is small enough
for the inaccuracy of the Gaussian approximation of the residual interference to
result in a discernible difference between the BER and $\tilde{P}_e$.
That said, even for this small block size, $\tilde{P}_e$ is an accurate approximation
of the BER in the absence of error propagation.

A few other features of Figs~\ref{fig:ZF_MIMO_M3P3}--\ref{fig:MMSE_MIMO_M3P4}
are worthy of note. First, the average performance of the direct and
DFT transmission schemes are essentially the same. This is to be expected
because the statistics of $\bH$ are unitarily invariant. 
Second, the increase in the diversity provided by the channel when using $P=4$~receiver antennas
rather than $P=3$ is clear from the different slopes of the BER curves
at high SNR. Finally, the performance advantage
of the optimized MMSE-BDFD scheme over the optimized ZF-BDFD scheme 
is significant in the case of $P=4$ receiver antennas
and is substantial in the case of $P=3$.  
The performance advantage of the optimized MMSE-BDFD scheme is due, in part,
to the fact the power  allocated to the first $M$ eigenmodes of
$\bH^H\bR_{vv}^{-1}\bH$ depends on  the corresponding eigenvalues. In particular,
weak eigenmodes might not be allocated any power at all.
In contrast, the optimized ZF-BDFD scheme allocates power uniformly over these
eigenmodes. The larger performance advantage of the optimized MMSE-BDFD scheme in the case
of $P=3$ is due to the larger probability of encountering a channel
 matrix such that $\bH^H\bR_{vv}^{-1}\bH$ 
does not have $M=3$ significant eigenvalues.

For reference, we have included the performance of a standard
orthogonal space-time block coding (OSTBC) scheme in
Figs~\ref{fig:ZF_MIMO_M3P3}--\ref{fig:MMSE_MIMO_M3P4}.
(Like the direct and DFT transmission schemes, OSTBC schemes were designed to
 be applied without knowledge of the channel at the transmitter.)
We have used the (symbol) rate $3/4$ code in~\cite{Ganesan_Stoica}
(which is a simplified version of that in~\cite{Tarokh}),
and hence in order to achieve a bit rate of 6 bpcu, a natural choice for
the underlying constellation is 256--QAM. (We assume that the channel is constant
for the four channel uses that are required to transmit the
codewords.) As expected, at high SNR, the OSTBC scheme provides
better BER performance than that direct transmission
(V-BLAST) scheme. However, the proposed precoder (which exploits
knowledge of the channel) provides substantially better performance
when $P=4$ receiver antennas are employed, and when $P=3$
and the MMSE-BDFD receiver is used.

When $P=3$ receiver antennas are employed and the ZF-BDFD is used,
the OSTBC scheme performs better than the optimized scheme at high SNRs.
This does not contradict the optimality of the proposed transceiver design, because the values
of $M$, $K$ and $P$, and the structure of the channel matrix, are different for the
OSTBC scheme.%
\footnote{In this example, the channel
matrix for the OSTBC scheme is $\bI_4\otimes\bH$,
where $\otimes$ denotes the Kronecker product and $\bH$ is the channel
matrix for the other schemes. The corresponding block
sizes are $P=12$, $K=12$, and $M=3$.}
The good performance of the OSTBC scheme at high SNRs is simply a manifestation
of the trade-off between error rate (achievable diversity) and symbol rate  in multiple
antenna fading channels without outer codes~\cite{V_Barry_ICC2002}.
(That trade-off is related to the fundamental
diversity-multiplexing  trade-off~\cite{ZhengTse_DivMux}.)
The symbol rate of the OSTBC scheme is significantly lower than that
of the proposed scheme.%
\footnote{In particular, in 4 consecutive channel
uses, the proposed scheme transmits $4M=12$ symbols, whereas
the OSTBC scheme transmits only 3~symbols.}
Hence, in the range of SNRs in which noise dominates the error performance,
the proposed scheme provides better performance than the OSTBC scheme, but
in the SNR range in which the channel condition dominates the error
performance, the OSTBC scheme provides better performance.
To illustrate that point, in Fig.~\ref{fig:ZF_MIMO_M3P3} we plotted
with unmarked curves the performance of the proposed ZF-BDFD scheme with 
a symbol rate of $M=2$ (as distinct from the scheme with $M=3$ described above).
In order to maintain a bit rate of 6~bpcu, the elements of
$\bs$ were taken, in an independent and equally-likely fashion,
from an 8--QAM constellations, and for consistency, the
SNR was defined to be $(p_0/3)/\sigma^2$.
Over the range of SNRs considered, the performance of the proposed
ZF-BDFD scheme with $M=2$ is substantially better than that of
the OSTBC scheme, with SNR gains of over 7~dB.

\begin{figure}
   \centering
      \resizebox{0.48\textwidth}{!}{\includegraphics{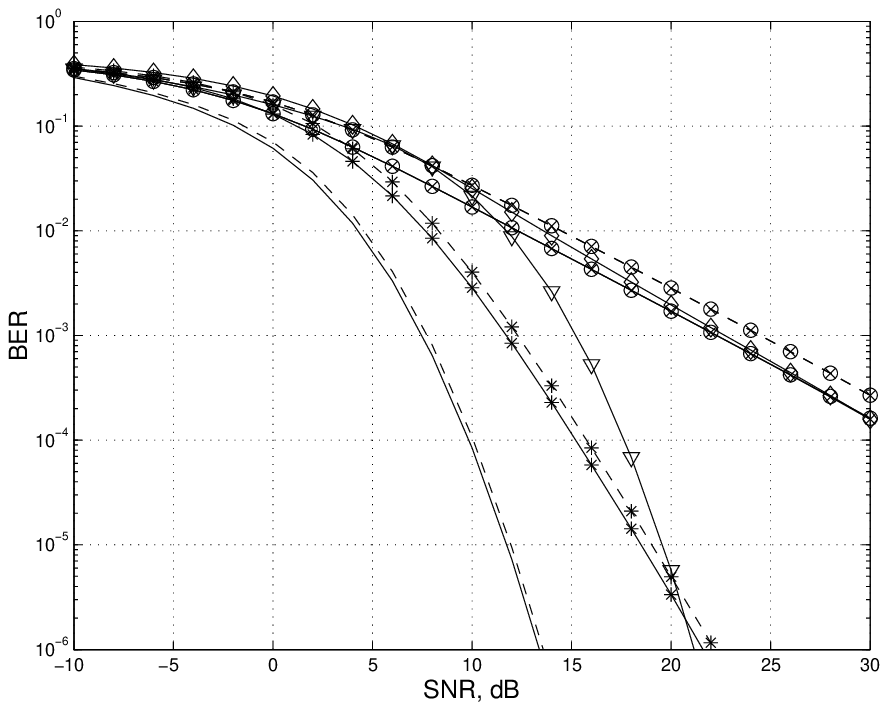}}
      \caption{Average BER performance of the ZF-BDFD for
the various precoders  and the linear ZF detector with its optimal
precoder in the narrowband multiple antenna scenario
 in Section~\ref{sec:MIMO} with $3$ transmitter antennas, $3$ receiver antennas
 and $M=3$ symbols per block.   The solid curves denote
performance achieved in the absence of error propagation, and the
dashed curves incorporate the effects of error propagation. 
Legend---$\ast$: optimized scheme,
$\bF_{\text{\textsc{opt-zf-bdfd}}}$;
$\circ$: direct, $\bF_{\text{\textsc{i}}}$; 
$\times$: DFT, $\bF_{\text{\textsc{dft}}}$;
$\diamond$: optimized linear ZF scheme, $\bF_{\text{\textsc{opt-zf-l}}}$;
$\triangledown$: OSTBC.
For later reference, the unmarked curves are for the optimized scheme with $M=2$.}
\label{fig:ZF_MIMO_M3P3}
\end{figure}

\begin{figure}
   \centering
      \resizebox{0.48\textwidth}{!}{\includegraphics{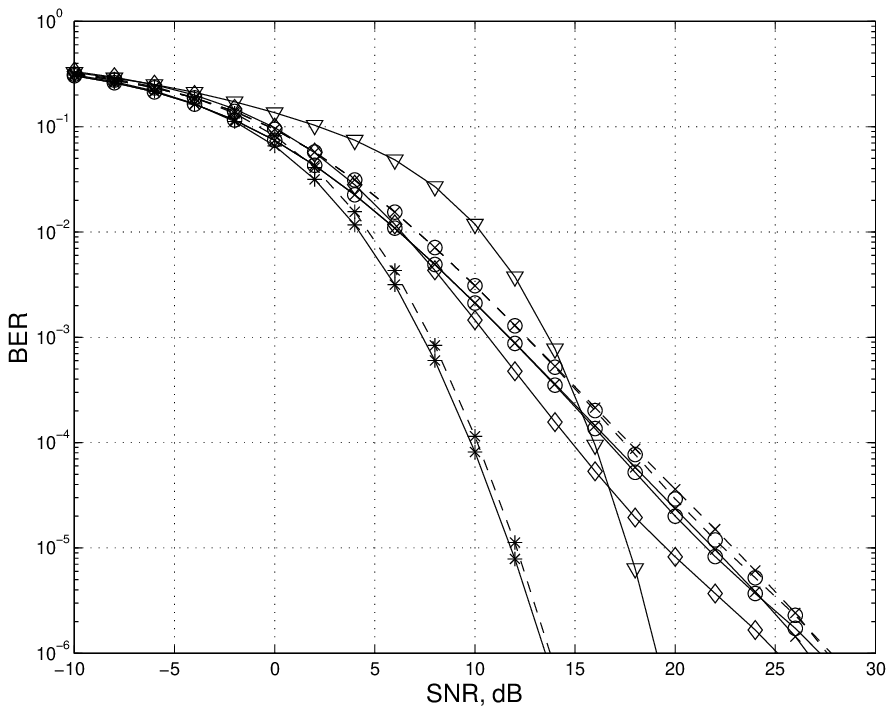}}
      \caption{Average BER performance of the ZF-BDFD for
the various precoders  and the linear ZF detector with its optimal
precoder in the narrowband multiple antenna scenario
 in Section~\ref{sec:MIMO} with 
 3 transmitter antennas and 4 receiver antennas.
The legend is the same as that in Figure~\ref{fig:ZF_MIMO_M3P3}.}
\label{fig:ZF_MIMO_M3P4}
\end{figure}

\begin{figure}
   \centering
      \resizebox{0.48\textwidth}{!}{\includegraphics{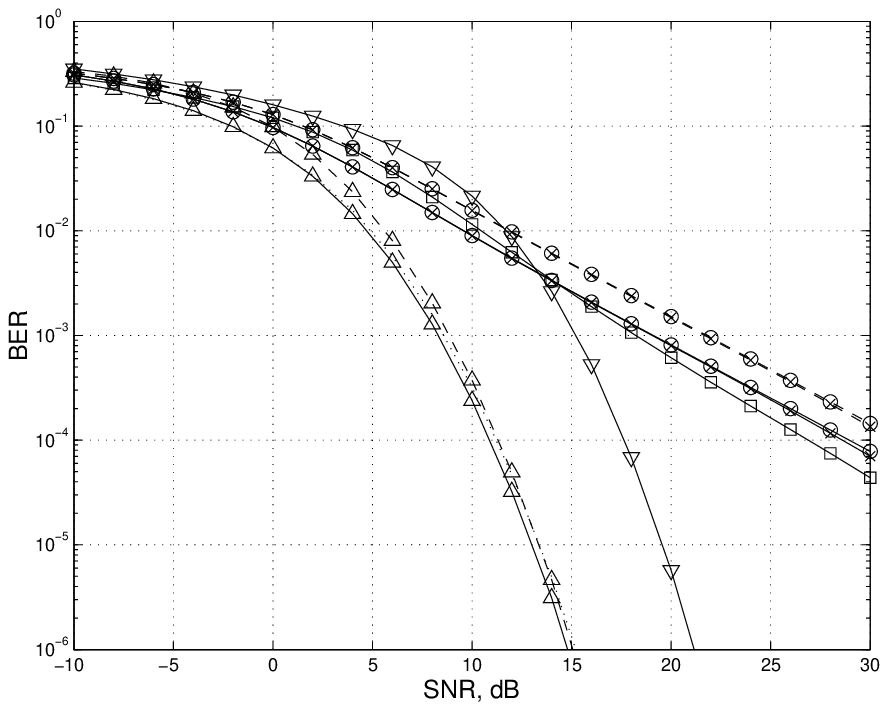}}
      \caption{Average BER performance of the MMSE-BDFD for
the various precoders  and the linear MMSE detector with its optimal
precoder in the narrowband multiple antenna scenario
 in Section~\ref{sec:MIMO} with 3 transmitter antennas and 3 receiver antennas.
  The solid curves denote
performance achieved in the absence of error propagation, and the
dashed curves incorporate the effects of error propagation. 
Legend---$\vartriangle$: optimized scheme,
$\bF_{\text{\textsc{opt-mmse-bdfd}}}$;
$\circ$: direct, $\bF_{\text{\textsc{i}}}$; 
$\times$: DFT, $\bF_{\text{\textsc{dft}}}$;
$\Box$: optimized linear ZF scheme, $\bF_{\text{\textsc{opt-mmse-l}}}$;
$\triangledown$: OSTBC.
The dotted curve denotes the lower bound on $\tilde{P}_e$ in \eqref{eq:BER_lowerbound_MMSE}.}
\label{fig:MMSE_MIMO_M3P3}
\end{figure}

\begin{figure}
   \centering
      \resizebox{0.48\textwidth}{!}{\includegraphics{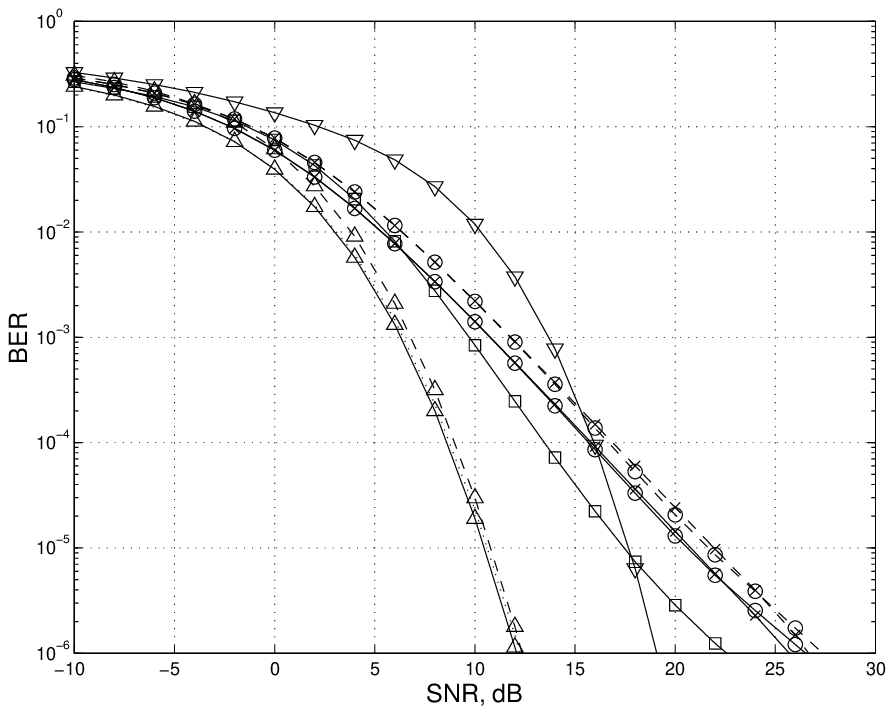}}
      \caption{Average BER performance of the MMSE-BDFD for
the various precoders  and the linear MMSE detector with its optimal
precoder in the narrowband multiple antenna scenario
 in Section~\ref{sec:MIMO}  with 3 transmitter antennas, 4 receiver antennas.
The legend is the same as that in Figure~\ref{fig:MMSE_MIMO_M3P3}.}
\label{fig:MMSE_MIMO_M3P4}
\end{figure}

\section{Conclusion}
\label{sec:conc}
In this paper, we have jointly designed the precoder and the feedback
 matrix of a block-by-block transmission scheme equipped with a
 zero-forcing or minimum mean-square error (MMSE) intra-block 
decision feedback detector (BDFD). The designs minimize the arithmetic
mean of the expected squared errors  at the decision point,
under the standard assumption that the previous symbols were
correctly detected. The covariance matrix of
 the minimized error is white, and hence the
proposed designs also minimize the (dominant components of
the) bit error rate of a uniformly bit-loaded transmission
system. In our simulations, the proposed systems performed significantly
better than  standard precoding systems, and retained their 
performance advantages in the presence of error propagation. 
In the case of the MMSE-BDFD, the proposed design also maximizes the
Gaussian mutual information. Since the MMSE-BDFD is a ``canonical'' 
receiver~\cite{cioffi,cioffi-forney-GDFE,Guess_Varanasi_MMSE_DFE},  
this suggests that by using the proposed transceiver design,
one can approach the capacity of the block transmission system
using (independent instances of) the same (Gaussian)
code for each element of the block.

\appendices
\section{Algorithm for Lemma~\ref{lem:JKZ}}
\label{sec:app_JKZ}
To state the algorithm succinctly, we make the following definitions:
$g=\bigl(\prod_{k=1}^M \gamma_k^2\bigr)^{1/M}$; $[\bS]_{\cdot k}$ 
denotes the $k$th column of
$\bS$ and $s_{\ell k}$ denotes its elements; $\bZ_k$ denotes the first
$k$ columns of $\bS$ and $\bZ_k^\perp$ denotes its orthogonal complement;
$\mathcal{P}_{\bA} = \bI - \bA(\bA^H\bA)^{-1}\bA^H$.
The recursion will be based on the $(M-k)\times (M-k)$ matrix
\begin{equation}
{\mathbf A}^{(k)}=\bigl(\boldsymbol{\Gamma}\mathbf{Z}_k^\perp\bigr)^H 
{\mathcal  P}_{(\boldsymbol{\Gamma}\mathbf{Z}_k^\perp)}
\boldsymbol{\Gamma}\mathbf{Z}_k^\perp.
\label{eq:A_k}
\end{equation}
For convenience, we assume that the elements of $\boldsymbol{\Gamma}$ are
arranged in non-increasing order. The algorithm proceeds as follows:

\begin{raggedright}
\begin{enumerate}
\item
Initialization: Set $k=1$. An explicit solution for
the first column of $\bS$ is 
$s_{11}=\sqrt{\frac{g-\gamma_M^2}{\gamma_1^2-\gamma_M^2}}$,
\newline
$s_{M1}=\sqrt{\frac{\gamma_1^2-g}{\gamma_1^2-\gamma_M^2}}$,
$\quad s_{\ell 1}=0$ for $\ell=2, 3, \cdots, M-1$.
\item
Construct $\bA^{(k)}$ in \eqref{eq:A_k} and its eigen decomposition,
${\mathbf A}^{(k)}={\mathbf V}^{(k)}{\boldsymbol
\Lambda}^{(k)} \bigl({\mathbf V}^{(k)}\bigr)^H$.
\item Set the $(k+1)$th column of $\bS$ to be 
$[\bS]_{\cdot k+1} ={\bZ}_k^\perp{\mathbf V}^{(k)}{\mathbf
y}^{(k)}$,
where
$y_1^{(k)}=\sqrt{\frac{g-
\lambda^{(k)}_{M-k}}{\lambda^{(k)}_1-\lambda^{(k)}_{M-k}}}$,
\newline
$y_{M-k}^{(k)}=\sqrt{\frac{\lambda^{(k)}_1-g}
{\lambda^{(k)}_1-\lambda^{(k)}_{M-k}}}$, 
$\quad y_{\ell}^{(k)}=0$ for
$\ell =2, 3, \cdots, M-k-1$.
\item
Increment $k$. If $k\leq M-2$ return to 2. Otherwise,
set $[\bS]_{\cdot M} ={\bZ}_{M-2}^\perp{\mathbf
V}^{(M-2)}{\mathbf y}^{(M-1)}$, where
$y_1^{(M-1)}=-\sqrt{\frac{g-\lambda^{(M-2)}_2}{\lambda^{(M-2)}_1-\lambda^{(M-2)}_2}}$,
$\quad y_2^{(M-1)}=\sqrt{\frac{\lambda^{(M-2)}_1-g}
{\lambda^{(M-2)}_1-\lambda^{(M-2)}_2}}$.
\end{enumerate}
\end{raggedright}

\section{Analytic Performance Comparisons}
\label{sec:app_BERcomp}

It was shown in Section~\ref{sec:BER}  that the precoders
designed in Section~\ref{sec:deriv} achieve the minimized value
of the lower bound on $\tilde{P}_e$; c.f.,~\eqref{Pelb} and \eqref{eq:BER_lowerbound_MMSE}.
Therefore, the relative BER performance of the optimized ZF-BDFD and
MMSE-BDFD systems in the absence of error propagation
 can be determined by simply comparing the
optimal values of the MSE, $\bar{e}^2 = \operatorname{tr}(\bR_{ee})/M$.
(A preliminary version of this appendix appeared in~\cite{Meng},
and related results on the MSEs of conventional decision feedback
equalizers appear in~\cite[Chapter~8]{BarryLeeMesserschmitt}.)

In order to ensure that the ZF systems exist, we will assume that
$\operatorname{rank}(\bH)\geq M$, and to simplify the comparisons, we
will also assume that the transmitted power $p_0$ is large enough
that $q=M$ in \eqref{MMSEFAI} for the MMSE-BDFD and $\ell=M$ 
in \eqref{eq:Fopt_linMMSE} for the linear MMSE detector.%
\footnote{The assumption  that $\operatorname{rank}(\bH)\geq M$
ensures that there is a threshold value for $p_0$ above which
$q=M$ and $\ell=M$.}
 Proposition~\ref{prop:mse_zf} states that the minimum value of the MSE
for a ZF-BDFD system is
\begin{equation*}
\bar{e}_{\text{\textsc{opt-zf-bdfd}}}^2= \frac{M}{p_0}\: |\tilde{\bLambda}_M|^{-1/M},
\end{equation*}
and Proposition~\ref{prop:mse_mmse} states that the minimum value
of the MSE  for an MMSE-BDFD system is 
\begin{equation*}
\bar{e}_{\text{\textsc{opt-mmse-bdfd}}}^2= 
\frac{M}{p_0+\operatorname{tr}(\tilde{\bLambda}_M^{-1})}\: 
|\tilde{\bLambda}|^{-1/M}.
\end{equation*}
Since $\tilde{\bLambda}_M$ is positive definite, 
$\bar{e}_{\text{\textsc{opt-mmse-bdfd}}}^2< 
\bar{e}_{\text{\textsc{opt-zf-bdfd}}}^2$, and hence, 
in the absence of error propagation, the optimized
MMSE-BDFD system will provide a lower BER than the optimized ZF-BDFD system.
While it is intuitively obvious that for a given precoder, the MMSE-BDFD
will provide a lower MSE than the ZF-BDFD, in the case of optimized 
precoders, this lower MSE leads directly to a lower BER.

The analysis of Section~\ref{sec:BER} remains valid for systems with
linear detectors, so long as the constraint $\bB=\boldsymbol{0}$ is
enforced. Therefore, we can compare the BER performance of an optimized
BDFD system with that of the system that is optimized for the
corresponding linear detector by simply comparing their minimum
MSEs.  The minimum MSE of a system with a linear ZF detector
is~\cite{dingy}
\begin{align*}
\bar{e}_{\text{\textsc{opt-zf-l}}}^2 &= \frac{1}{Mp_0}\: 
\bigl(\operatorname{tr}(\tilde{\bLambda}_M^{-1/2})\bigr)^2,\\
&\geq \frac{M}{p_0}\: |\tilde{\bLambda}_M|^{-1/M}
= \bar{e}_{\text{\textsc{opt-zf-bdfd}}}^2,
\end{align*}
where we have used the trace-determinant inequality~\eqref{eq:AGMI}.
Therefore, in the absence of error propagation 
the optimized system for the ZF-BDFD will provide
a lower BER than the optimized system for the linear ZF detector.
Similarly, the minimum MSE of a system with a linear MMSE detector
is~\cite{chan,palomar}
\begin{align*}
\,&\bar{e}_{\text{\textsc{opt-mmse-l}}}^2  \\ 
&\qquad = 
\frac{1}{M\bigl(p_0+\operatorname{tr}(\tilde{\bLambda}_M^{-1})\bigr) - 
\bigl(\operatorname{tr}(\tilde{\bLambda}_M^{-1/2})\bigr)^2}\:
\bigl(\operatorname{tr}(\tilde{\bLambda}_M^{-1/2})\bigr)^2,\\
&\qquad > \frac{M}{p_0+\operatorname{tr}(\tilde{\bLambda}_M^{-1})}\: 
|\tilde{\bLambda}_M|^{-1/M}
= \bar{e}_{\text{\textsc{opt-mmse-bdfd}}}^2,
\end{align*}
and hence the optimized system for the MMSE-BDFD provides a
lower BER than the optimized system for the linear MMSE detector.
As observed in~\cite{chan}, 
$\bar{e}_{\text{\textsc{opt-mmse-l}}}^2  \leq
\bar{e}_{\text{\textsc{opt-zf-l}}}^2$, and hence
the optimized system for the linear MMSE detector provides a lower
BER than the optimized system for the linear ZF detector.

\section*{Acknowledgment}
The authors gratefully acknowledge the assistance of 
Scarlett Chan of McMaster University in the preparation of
the simulation results reported in Section~\ref{sec:FIR},
and that of Qian Meng of McMaster University in the preparation
of Appendix~\ref{sec:app_BERcomp}.

\end{document}